%% file: 00_abstract.tex
\documentclass[acmsmall,screen]{acmart}

\usepackage{amsmath,amsfonts}
\usepackage[ruled,vlined,linesnumbered]{algorithm2e}
\usepackage{array}
\usepackage{textcomp}
\usepackage{stfloats}
\usepackage{url}
\usepackage{verbatim}
\usepackage{romannum}
\usepackage{graphicx}
\usepackage{xspace}
\usepackage{xcolor}
\usepackage{hyperref}
\usepackage[export]{adjustbox}
\usepackage{subcaption}
\usepackage{caption}
\usepackage{listings}
\usepackage[noend]{algpseudocode}
\usepackage{balance}
\usepackage[T1]{fontenc}   

\def\BibTeX{{\rm B\kern-.05em{\sc i\kern-.025em b}\kern-.08em
    T\kern-.1667em\lower.7ex\hbox{E}\kern-.125emX}}

\newcommand{\ExMalloc}{\textit{Exgen-Malloc}\xspace}

\begin{document}

\setcopyright{none}
\settopmatter{printacmref=false} 
\renewcommand\footnotetextcopyrightpermission[1]{} 

\author{Ruihao Li}
\email{liruihao@utexas.edu}
\affiliation{%
  \institution{The University of Texas at Austin}
  \city{Austin}
  \state{Texas}
  \country{USA}
}

\author{Lizy K. John}
\email{ljohn@ece.utexas.edu}
\affiliation{%
  \institution{The University of Texas at Austin}
  \city{Austin}
  \state{Texas}
  \country{USA}
}

\author{Neeraja J. Yadwadkar}
\email{neeraja@austin.utexas.edu }
\affiliation{%
  \institution{The University of Texas at Austin}
  \city{Austin}
  \state{Texas}
  \country{USA}
}

\title{Old is Gold: Optimizing Single-threaded Applications with Exgen-Malloc}
\begin{abstract}

Memory allocators hide beneath nearly every application stack, yet their performance footprint extends far beyond their code size.
Even small inefficiencies in the allocators ripple through caches and the rest of the memory hierarchy, collectively imposing what operators often call a ``datacenter tax''.
At hyperscale, even a 1\% improvement in allocator efficiency can unlock millions of dollars in savings and measurable reductions in datacenter energy consumption.
Modern memory allocators are designed to optimize allocation speed and memory fragmentation in multi-threaded environments, relying on complex metadata and control logic to achieve high performance. 
However, the overhead introduced by this complexity prompts a reevaluation of allocator design. 
Notably, such overhead can be avoided in single-threaded scenarios, which continue to be widely used across diverse application domains. \par

In this paper, we introduce \ExMalloc, a memory allocator purpose-built for single-threaded applications.
By specializing for single-threaded execution, \ExMalloc eliminates unnecessary metadata, simplifies the control flow, thereby reducing overhead and improving allocation efficiency.
Its core design features include a centralized heap, a single free-block list, and a balanced strategy for memory commitment and relocation.
Additionally, \ExMalloc incorporates design principles in modern multi-threaded allocators, which do not exist in legacy single-threaded allocators such as dlmalloc.
We evaluate \ExMalloc on two Intel Xeon platforms. 
Across both systems, \ExMalloc achieves a speedup of $1.17\times$, $1.10\times$, and $1.93\times$ over dlmalloc on SPEC CPU2017, redis-benchmark, and mimalloc-bench, respectively.
In addition to performance, \ExMalloc achieves $6.2\%$, $0.1\%$, and $25.2\%$ memory savings over mimalloc on SPEC CPU2017, redis-benchmark, and mimalloc-bench, respectively.
\par
\end{abstract}

\maketitle

\pagenumbering{arabic}
\input{content}


\bibliographystyle{ACM-Reference-Format}
\bibliography{refs}

\end{document}

%% file: content.tex
\section{Introduction}
\label{section:introduction}
\input{01_introduction}

\section{Background and Motivation}
\label{section:background}
\input{02_background}

\section{Design of \ExMalloc}
\label{section:design}
\input{03_00_design}

\section{Implementation and Usage of \ExMalloc}
\label{section:using}
\input{04_using}

\section{Evaluation}
\label{section:prototype}
\input{05_00_evaluation}

\section{Conclusion}
\label{section:conclusion}
\input{06_conclusion}

%% file: 01_introduction.tex
\begin{figure}[t]
    \centerline{\includegraphics[width=0.8\columnwidth, trim = 19mm 45mm 80mm 37mm, page=7, clip=true]{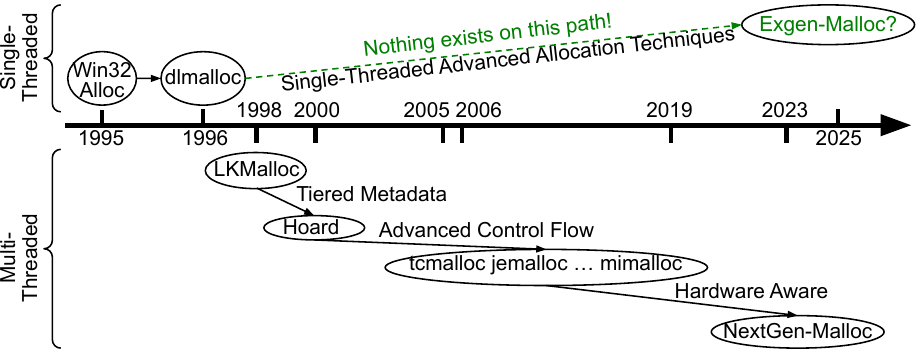}}
	\caption{Timeline of memory allocators (for C/C++). Since LKMalloc, efforts have primarily focused on multi-threaded allocators, leaving single-threaded allocators largely overlooked. }
	\label{fig_timeline}
\end{figure}

With the increasing prevalence of multi-threaded applications that exploit parallelism for performance gains, modern memory allocators are typically architected to support concurrent allocation and deallocation requests across multiple threads, minimizing contention and maximizing scalability.
To meet the demands of multi-threaded applications, memory allocators have themselves shifted from single-threaded to multi-threaded library designs.
Fig.~\ref{fig_timeline} shows the evolution of memory allocators over time.
During the mid-1990s, single-threaded memory allocators, such as the Win32 allocator and dlmalloc~\cite{richter1995advanced, lea1996memory, berger2002reconsidering} (also referred to as Windows XP memory allocator and Lea allocator), 
were predominant, reflecting the widespread use of single-core processors.
By the late 1990s and early 2000s, with the rise of multi-core processors, LKMalloc~\cite{larson1998memory} became one of the first multi-threaded memory allocators. 
Advancements introduced techniques like tiered metadata and better metadata management with improved control flow, culminating in the development of allocators such as Hoard~\cite{berger2000hoard}, tcmalloc~\cite{tcmalloc} from Google, jemalloc~\cite{evans2006scalable} widely used at Meta, and mimalloc~\cite{leijen2019mimalloc} from Microsoft. \par

These modern memory allocators are complex pieces of code that need to excel at many tasks, including (a) performing fast allocation and deallocation of objects, (b) handling objects of various sizes efficiently with minimal fragmentation, (c) returning freed space quickly so that further allocations can be made, (d) being scalable to many threads and cores, and so on.
To achieve these tasks, modern memory allocators rely on complex metadata~\cite{berger2000hoard, evans2006scalable, leijen2019mimalloc, lietar2019snmalloc, hunter2021beyond}. \par

While most allocator designs in recent years have primarily targeted multi-threaded applications, single-threaded workloads have largely been overlooked. 
Nevertheless, single-threaded applications remain prevalent across a wide range of domains due to their simplicity, ease of testing, and reduced debugging complexity~\cite{wang2023memento, karandikar2023cdpu, redis}. 
Common examples include serverless functions~\cite{wang2023memento}, many data compression algorithms~\cite{karandikar2023cdpu}, and Redis, a widely adopted in-memory database that operates in a single-threaded manner~\cite{redis}.
Additionally, single-threaded applications are always popular in mobile and embedded devices, where energy constraints make lightweight and low-overhead execution essential. 
Modern multi-threaded allocators introduce significant complexity to ensure thread safety, often at the cost of performance in single-threaded scenarios. 
Studies have shown that these allocators can degrade performance due to cache pollution, as allocator metadata displaces cache lines used by application data~\cite{li2023nextgen}. 
In contrast, single-threaded applications place fewer demands on memory allocators and do not require synchronization mechanisms. 
However, the widespread use of allocators optimized for multi-threaded workloads -- even in single-threaded contexts -- introduces unnecessary overheads such as synchronization primitives (e.g., locks) and additional metadata management. 
These overheads, while essential for thread safety in multi-threaded environments, become superfluous in single-threaded settings and can lead to measurable performance degradation. \par

We argue that if a program is statically known to be single-threaded, a specialized memory allocator optimized for single-threaded execution can be safely selected at link time. 
This approach is practical, as determining the threading model of a program--whether single-threaded or multi-threaded--is feasible during compile-time or link-time analysis~\cite{tseng2014cdtt} (details in \S~\ref{section:using}).
While legacy single-threaded allocators such as dlmalloc~\cite{lea1996memory} naturally avoid synchronization and metadata overheads, modern systems demand support for complex allocation patterns, nuanced locality behaviors, and diverse object-size distributions~\cite{zhou2024characterizing}.
To this end, we present \ExMalloc, a memory allocator designed specifically for single-threaded applications that retains a single-threaded design while integrating key principles from modern multi-threaded allocators.
Specifically, \ExMalloc leverages single-thread–optimized techniques -- namely, a centralized heap, a single free-block list, and a balanced strategy for memory commitment and relocation -- to simplify complex hierarchical metadata management and eliminate the locks and atomic operations required by multi-threaded allocators.
Meanwhile, \ExMalloc incorporates optimizations inspired by contemporary multi-threaded allocators, such as aggregated metadata for fast indexing and efficient reuse of freed blocks (details in \S~\ref{section:design}). \par

\begin{figure}[t]
    \centerline{\includegraphics[width=0.6\columnwidth, trim = 8mm 8mm 8mm 8mm, page=1, clip=true]{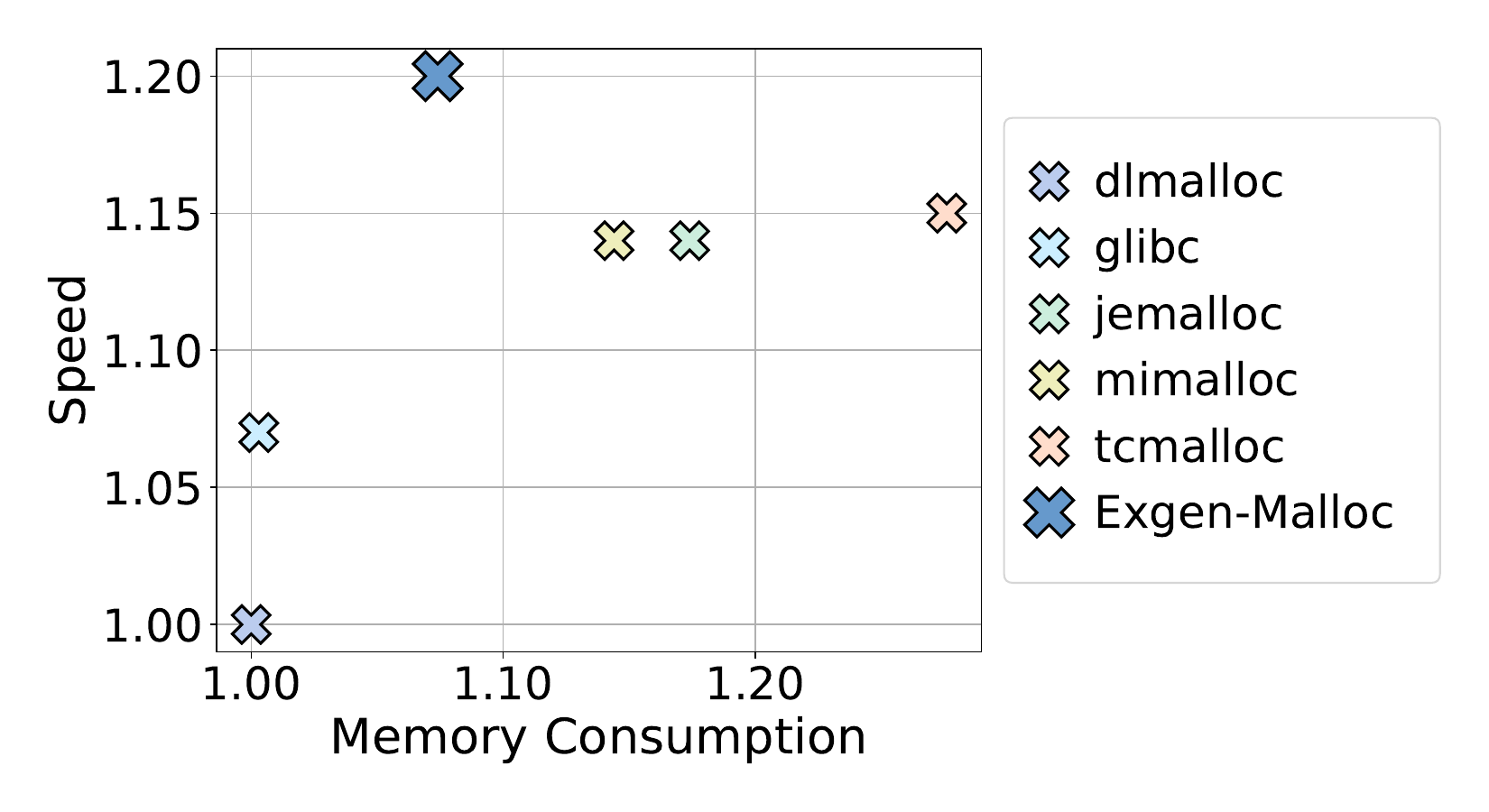}}
	\caption{Performance and memory efficiency comparison of \ExMalloc against the legacy single-threaded allocator (dlmalloc), the default glibc allocator, and modern multi-threaded allocators (jemalloc, tcmalloc, and mimalloc). \ExMalloc achieves higher speedups and lower memory consumption than state-of-the-art multi-threaded allocators.}
	\label{fig_overview}
\end{figure}

By combining mechanisms specifically tailored to single-threaded execution and optimization techniques from modern multi-threaded allocators, \ExMalloc achieves both higher performance and lower memory consumption than state-of-the-art multi-threaded allocators (jemalloc, tcmalloc, and mimalloc, as shown in Fig.~\ref{fig_overview}).
We rigorously evaluate the performance of memory consumption of \ExMalloc on two Intel Xeon platforms using SPEC CPU2017, redis-benchmark, and mimalloc-bench, these three benchmark suites. 
We compare \ExMalloc against the single-threaded allocator dlmalloc~\cite{lea1996memory}, the glibc allocator, and three state-of-the-art multi-threaded allocators--jemalloc~\cite{evans2006scalable}, tcmalloc~\cite{tcmalloc}, and mimalloc~\cite{leijen2019mimalloc}.
Across the two evaluated systems, \ExMalloc achieves a geometric mean speedup of $1.17\times$ over dlmalloc and $1.05\times$ over mimalloc on the SPEC CPU2017 benchmark suite. 
On redis-benchmark, \ExMalloc delivers $1.10\times$ and $1.01\times$ performance improvements over dlmalloc and mimalloc, respectively, and on mimalloc-bench, it further improves performance by $1.93\times$ and $1.02\times$.
Beyond performance, \ExMalloc enhances memory efficiency, reducing memory usage by $6.2\%$, $0.1\%$, and $25.2\%$ compared to mimalloc on SPEC CPU2017, redis-benchmark, and mimalloc-bench, respectively.
\par

%% file: 02_background.tex
We first provide an overview of memory allocators (\S~\ref{section_02_01}) and discuss the evolution from single-threaded to multi-threaded designs (\S~\ref{section_02_02}). 
We then highlight the limitations of universally applying multi-threaded allocators to all applications (\S~\ref{section_02_03}) and motivate the need for single-threaded allocators tailored to single-threaded programs (\S~\ref{section_02_04}).

\subsection{Background: Memory Allocators}
\label{section_02_01}
Memory allocators manage heap memory within the virtual address space of each process, typically via libraries such as mimalloc/tcmalloc, using functions like $malloc()$ and $free()$ (in C/C++). 
To track heap usage, allocators maintain internal bookkeeping structures, called allocator metadata. 
While programs can request memory directly from the operating system through system calls such as $mmap()$, each invocation incurs relatively high latency, typically on the order of microseconds.
State-of-the-art allocators (as illustrated in Algorithm~\ref{alg_malloc}) mitigate this overhead by preallocating large memory regions from the OS and managing allocations within these regions.
This design allows over 90\% of allocation requests to be served entirely in user space (malloc\_fast in Algorithm~\ref{alg_malloc}) with nanosecond-level latency~\cite{kanev2017mallacc, hunter2021beyond, zhou2024characterizing}, albeit at the cost of potential internal fragmentation within the preallocated large memory regions. \par

\textit{Balancing allocation speed and memory fragmentation remains a longstanding challenge in memory allocator design, drawing continued attention from both software and hardware communities~\cite{berger2000hoard, leijen2019mimalloc, hunter2021beyond, kanev2017mallacc, wang2023memento, zhou2024characterizing, apostolakis2025necro, zhou2023memperf, reitz2024starmalloc, aigner2015fast, lietar2019snmalloc, maas2021adaptive, yang2023numalloc, evans2006scalable}.}
Most allocators, for example, rely on linked-list-based data structures to track free blocks of varying sizes: coarser-grained blocks enable faster lookup and allocation but exacerbate fragmentation, whereas finer-grained blocks reduce fragmentation at the cost of longer allocation latency.

\lstset{
  basicstyle=\ttfamily\scriptsize,
  keywordstyle=\color{blue},
  commentstyle=\color{gray},
  stringstyle=\color{red},
  numbers=left,
  numberstyle=\tiny,
  breaklines=true,
  frame=single,
  captionpos=b
}

\begin{figure}[htbp]
\centering
\begin{minipage}{0.8\linewidth}
\centering
\lstset{language=C}
\begin{lstlisting}[caption={Pseudo code of malloc().}, label={alg_malloc}]
void* malloc(size_t size) {
    void *block = malloc_fast(size);
    if (block == NULL) return malloc_generic(size);
    return block;
}
void* malloc_fast(size_t size) {
    return block from OS pre-allocated memory;
}
void* malloc_generic(size_t size) {
    OS pre-allocate memory (mmap() syscall);
    return malloc_fast(size);
}
\end{lstlisting}
\end{minipage}
\end{figure}

\subsection{Transition from Single-threaded to Multi-threaded Allocators}
\label{section_02_02}
Modern multi-threaded applications typically issue memory allocation requests concurrently from multiple threads.
This has led to the transition from single-threaded memory allocators like dlmalloc~\cite{lea1996memory}, to multi-threaded allocators like LKMalloc~\cite{larson1998memory} and Hoard~\cite{berger2000hoard}, a trend that began in the late 1990s.
However, multi-threaded allocators are more complex than single-threaded allocators, introducing additional challenges in balancing allocation speed and memory fragmentation.

To enable concurrent allocation, multi-threaded allocators typically adopt one of the following two strategies: (a) assigning each thread a private memory pool, or (b) using a large shared memory region accessible by all threads. Each approach offers distinct benefits and trade-offs.
Allocating a dedicated memory pool for each thread can lead to a linear increase in memory consumption relative to the number of threads, a phenomenon known as memory \textit{blowup}~\cite{berger2000hoard}.
Conversely, using a centralized memory pool shared among all threads can introduce significant cross-core synchronization overhead due to lock contention. \par

\begin{figure}[t]
    \centerline{\includegraphics[width=0.7\columnwidth, trim = 1mm 13mm 51mm 3mm, page=1, clip=true]{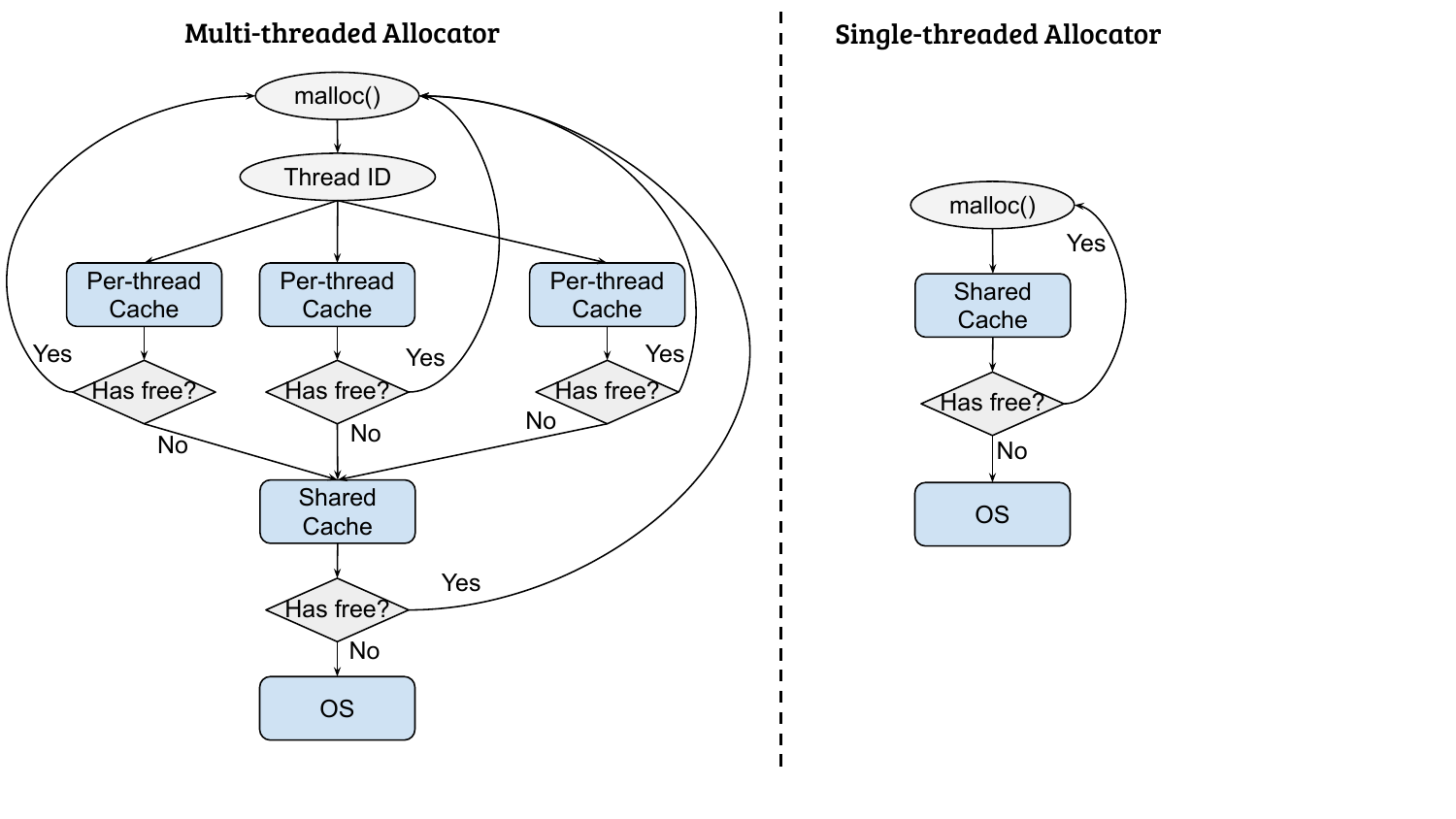}}
	\caption{Multi-threaded Allocator vs Single-threaded Allocator. The single-threaded allocator uses single-layer metadata and simplifies the control logic.}
	\label{fig_overview}
\end{figure}

To balance the trade-offs between per-thread and centralized memory pools, modern allocators leverage a hybrid approach. 
They utilize thread-local caches for small, fast allocations and maintain a centralized cross-thread shared memory pool to balance memory consumption across threads.
As the example shown in Fig.~\ref{fig_overview} (left side)\footnote{Fig.~\ref{fig_overview} represents a generalized abstraction of modern memory allocators, though specific implementations may vary. For instance, tcmalloc employs a three-tier hierarchy and uses per-core instead of per-thread cache~\cite{hunter2021beyond, zhou2024characterizing}.}, each $malloc()$ request first attempts to allocate memory from the thread-local cache. 
If the free memory space is insufficient, the allocator accesses the central memory pool to fulfill the allocation request. 
The per-thread cache and the shared cache synchronize periodically to ensure a balanced distribution of free memory among threads. 
While this hybrid approach mitigates the \textit{blowup} problem through tiered metadata, it can also introduce (a) additional cache pollution and (b) synchronization overhead~\cite{li2023nextgen, leijen2019mimalloc, lietar2019snmalloc, hunter2021beyond, aigner2015fast, kanev2017mallacc}. \par

\noindent{\textbf{Cache pollution:}}
Accessing the metadata can pollute the CPU caches by evicting cache lines containing user data~\cite{li2023nextgen}. 
In multi-threaded allocators, one allocation request can result in: (a) accessing the thread-local cache or (b) accessing the shared cache if the thread-local cache is exhausted, and potentially the thread-local cache of other threads for additional free space.
Cache pollution becomes an even greater challenge in multi-threaded applications, as allocator metadata can pollute cache lines of other cores, leading to the false-sharing issue, where independent variables modified by different threads reside in the same cache line~\cite{berger2000hoard}. \par

\noindent{\textbf{Synchronization:}}
Modern allocators use intensive synchronization primitives (e.g., locks) to secure the metadata coherency (for both thread-local and shared cache). 
Cross-core synchronization can downgrade the system performance since the communication overhead increases with the number of cores in the system~\cite{asgharzadeh2022free, wu2021virtual, wang2016caf}. \par

\subsection{Multi-threaded Allocators in Single-threaded Programs}
\label{section_02_03}
When multi-threaded memory allocators are used for single-threaded programs, they impose unnecessary performance overhead.
The potential overhead includes:
\begin{itemize}
    \item Additional tiered metadata data structure. As the \textit{blowup} issue does not exist in single-threaded programs, a single metadata layer is sufficient to manage the available memory blocks.
    \item Additional synchronization and control logic. As single-tiered metadata is enough for single-threaded programs, allocators do not need complex control logic, e.g., mapping each allocation request to each thread-local cache.
\end{itemize}

\begin{figure}[t]
    \centerline{\includegraphics[width=0.8\columnwidth, trim = 4mm 4mm 3mm 4mm, page=1, clip=true]{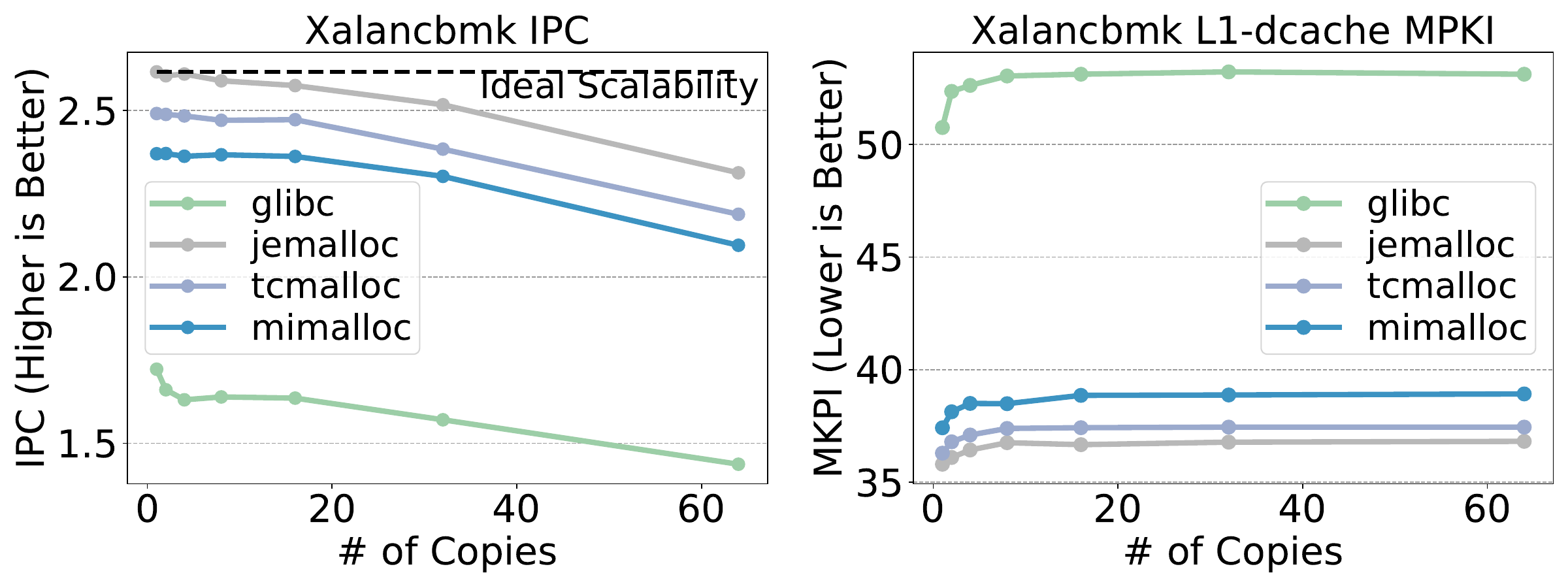}}
	\caption{Scalability of different memory allocators for \textit{xalancbmk}. 
    As the number of copies increases, the IPC decreases due to higher overhead caused by an increase in L1-dcache MPKI (misses per thousand instructions).
    }
	\label{fig_malloc_scalability}
\end{figure}

Fig.~\ref{fig_malloc_scalability} gives an example of the scalability of using multi-threaded allocators for \textit{single-threaded} application \textit{xalancbmk}, which applies XSLT transformations to XML documents using the Xalan-C++ processor. 
\textit{Xalancbmk} is a representative workload from SPEC CPU2017~\cite{spec}, which is broadly used for evaluating processors and compilers~\cite{panda2018wait, lee2018reconciling, mukherjee2020dataflow, livinskii2020random, poesia2020dynamic, weber2025relaxing} as well as memory allocators~\cite{aigner2015fast, kanev2017mallacc, lietar2019snmalloc}.
The number of copies in SPEC benchmarks is the count of concurrent instances (processes) of the same program, used to characterize the maximum throughput of a machine~\cite{spec}. \par

We observe the program performance drops as the number of copies increases: the per-core instruction per cycle (IPC) drops with each additional copy (left panel of Fig.~\ref{fig_malloc_scalability}). 
Because IPC measures how many instructions a core retires per clock, a lower value means the core completes less work per unit time--i.e., lower performance. 
To diagnose the cause, we profiled private L1 data cache (L1-dcache) misses (right panel of Fig.~\ref{fig_malloc_scalability}). 
In an ideal setup where each copy runs on its own core and allocator state is private, L1-dcache misses should remain roughly flat as copies scale. 
Instead, with multi-threaded allocators, we see L1-dcache misses rise with copy count, consistent with frequent accesses to shared allocator metadata (e.g., global free lists or page/arena headers). 
This sharing induces coherence traffic and cache line ping-pong, which reduces IPC and explains the observed slowdown. \par

\subsection{A Case for Single-Threaded Allocators}
\label{section_02_04}
The performance overhead of multi-threaded allocators suggests they may not be suitable for all applications.
Notably, after LKMalloc~\cite{larson1998memory} and Hoard in the early 2000s~\cite{berger2000hoard}, research has predominantly focused on multi-threaded scenarios~\cite{berger2002reconsidering, evans2006scalable, pheatt2008intel, ptmalloc, tcmalloc, leijen2019mimalloc, lietar2019snmalloc, hunter2021beyond, zhou2024characterizing}. 
However, single-threaded applications remain popular as they are often simpler to develop, test, and debug, making them an attractive choice for developers who prioritize simplicity and reliability. 
Moreover, the overhead associated with thread management in multi-threaded applications can sometimes negate the performance benefits for certain types of workloads. 
A wide range of performance-critical workloads remain single-threaded, including \textit{serverless} functions~\cite{wang2023memento}, \textit{data compression} algorithms~\cite{karandikar2023cdpu}, and in-memory databases such as \textit{Redis}~\cite{redis}. 
This highlights the pressing need for memory allocators specifically optimized to maximize efficiency in single-threaded applications. \par

Leveraging a memory allocator designed specifically for single-threaded applications can eliminate the overhead inherent in multi-threaded allocators (\S~\ref{section_02_03}). 
By simplifying internal logic, such allocators are inherently more lightweight—an advantage for datacenter workloads, which often rely on static linking for compatibility and deployment convenience, but at the cost of significantly larger program binaries~\cite{kanev2015profiling, sriraman2019softsku}. 
Beyond performance, single-threaded allocators also align better with environmental and budget constraints. 
In energy-constrained edge environments, single-threaded applications are preferred for their simplicity and energy efficiency~\cite{crotty2021case}, while their cost advantages~\cite{mcsherry2015scalability, balduini2018cost} make them particularly appealing for large-scale cloud platforms, where public clouds commonly provide single-core virtual machine options~\cite{AWS}. \par

%% file: 03_00_design.tex
In this paper, we argue that it is time to resurrect single-threaded memory allocators, especially given the continued popularity of single-threaded applications in the cloud and the datacenters~\cite{wang2023memento, karandikar2023cdpu, redis, kim2019functionbench, gan2019open}. 
By eliminating unnecessary metadata and simplifying control logic, while retaining key techniques from modern multi-threaded allocators, \ExMalloc can achieve high efficiency in single-threaded applications. \par

We introduce techniques tailored to optimize performance in single-threaded environments:
\begin{itemize}
\item A centralized heap that eliminates unnecessary hierarchical layers for single-threaded applications (\S~\ref{section_heap_layout}).
\item A per-page single free-block list for fast indexing and improved data locality (\S~\ref{section_no_local_free}).
\item Balancing memory commitment and reclamation to optimize allocation speed and overall memory efficiency (\S~\ref{section_commit}).
\end{itemize}

Additionally, we cannot rely solely on legacy single-threaded allocators such as dlmalloc~\cite{lea1996memory}, as contemporary single-threaded applications depict increasingly complex data locality, allocation patterns, and object size distributions characteristic of modern software~\cite{zhou2024characterizing}.
To meet these demands, \ExMalloc also incorporates design principles inspired by modern multi-threaded allocators, including:
\begin{itemize}
    \item Aggregated metadata for efficient indexing and rapid reuse of freed memory blocks, compact metadata structures to accelerate allocation speed, fine-grained block-size classes to reduce internal fragmentation, and inline functions to minimize the overhead of allocation function calls (\S~\ref{section_other}).
\end{itemize}

\input{03_01}

\input{03_02}
\input{03_03}
\input{03_04}

%% file: 03_01.tex
\begin{figure}[t]
    \centerline{\includegraphics[width=0.9\columnwidth, trim = 15mm 40mm 7mm 5mm, page=6, clip=true]{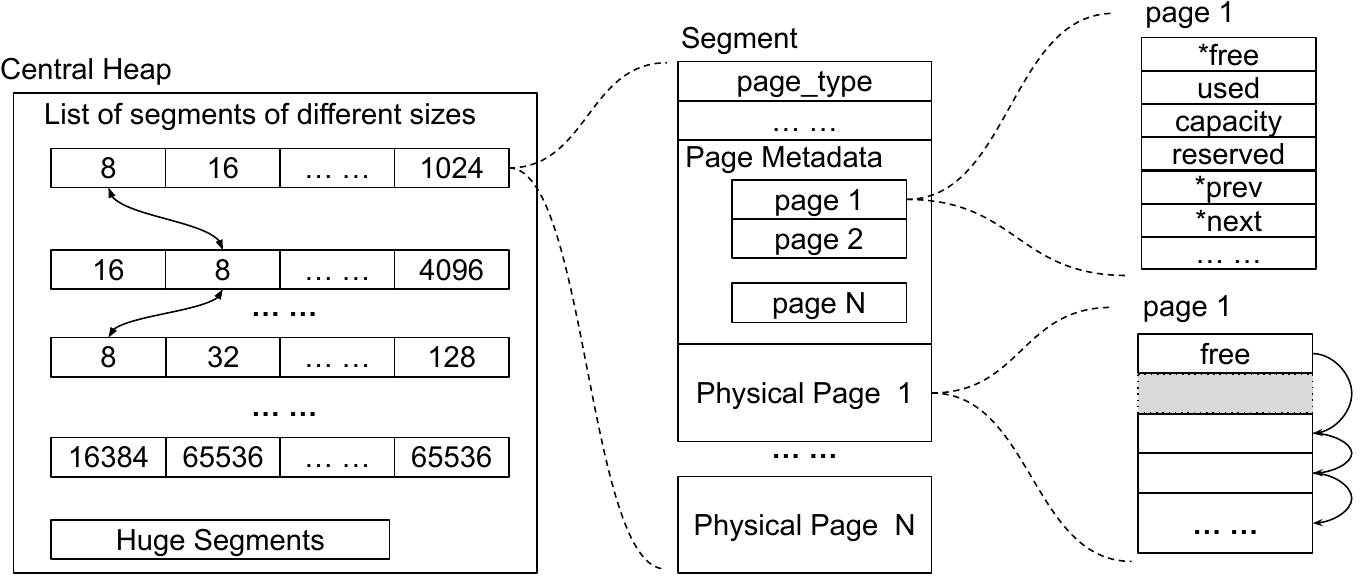}}
	\caption{Metadata layout of \ExMalloc. \ExMalloc employs a central heap composed of multiple segments, each of which is further divided into pages.}
	\label{fig_heap_layout}
\end{figure}

\subsection{A Centralized Heap}
\label{section_heap_layout}
\ExMalloc adopts a single centralized heap that manages segments across all size classes, simplifying metadata organization.
Fig.~\ref{fig_heap_layout} illustrates the metadata and memory layout of \ExMalloc.
In contrast to modern multi-threaded allocators such as tcmalloc and mimalloc, which rely on per-thread heaps and cross-thread coordination, \ExMalloc uses a unified heap design tailored specifically for single-threaded applications. \par

In the central heap, segments are organized into block-size classes (e.g., a size class of 8 represents 8-byte blocks; allocation requests from 1 to 8 bytes are aligned to 8 bytes and served with an 8-byte block). 
Each segment functions as a second-level metadata unit.
We use a 4 MiB segment size (more detailed in \S~\ref{section_commit}) to balance allocation speed and memory consumption.
Each segment is divided into multiple pages according to its page type. \ExMalloc defines four distinct page types. 
For the smallest page size, each segment contains 64 pages; for the medium page size, 8 pages; and for the largest size, a single page per segment.
Allocations that exceed the standard segment capacity are treated as huge segments, where the segment size equals the requested object size, as illustrated in Fig.~\ref{fig_heap_layout}. \par

The heap serves as a logical abstraction that contains no user data, whereas a segment is a hybrid structure encompassing both metadata -- including the page type, segment size, number of reserved and used pages, per-page metadata, and offsets to physical pages -- and the physical memory it occupies.
The metadata of each page tracks key attributes such as block size, total capacity, and the counts of allocated and freed blocks. 
The page metadata also maintains a free-block list pointer that directly references the address space of the page.
Additionally, each page stores pointers to its predecessor and successor within the same page type, forming a doubly linked list for each block-size class.
This organization enables fast indexing of pages within a segment, allowing each segment to host pages of different sizes while reducing both fragmentation and metadata overhead.
Inside a physical page, aggregated metadata (more details in \S~\ref{section_other}) and a single free-block list (more details in \S~\ref{section_no_local_free}) enable fast indexing with low memory fragmentation. \par

\noindent\textbf{Simplified control logic:}
While multi-threaded allocators require additional mechanisms for inter-thread coordination, increasing metadata complexity, a centralized heap suffices for single-threaded applications that do not perform cross-thread memory operations.
\ExMalloc leverages its single-threaded design to adopt a centralized heap, thereby eliminating remote metadata coordination and reducing overall metadata overhead.

%% file: 03_02.tex
\begin{figure*}[t]
	\begin{minipage}[t]{0.54\textwidth}
		\centering
	\includegraphics[width=\textwidth, trim = 5mm 0mm 58mm 0mm, clip=true, page=4]{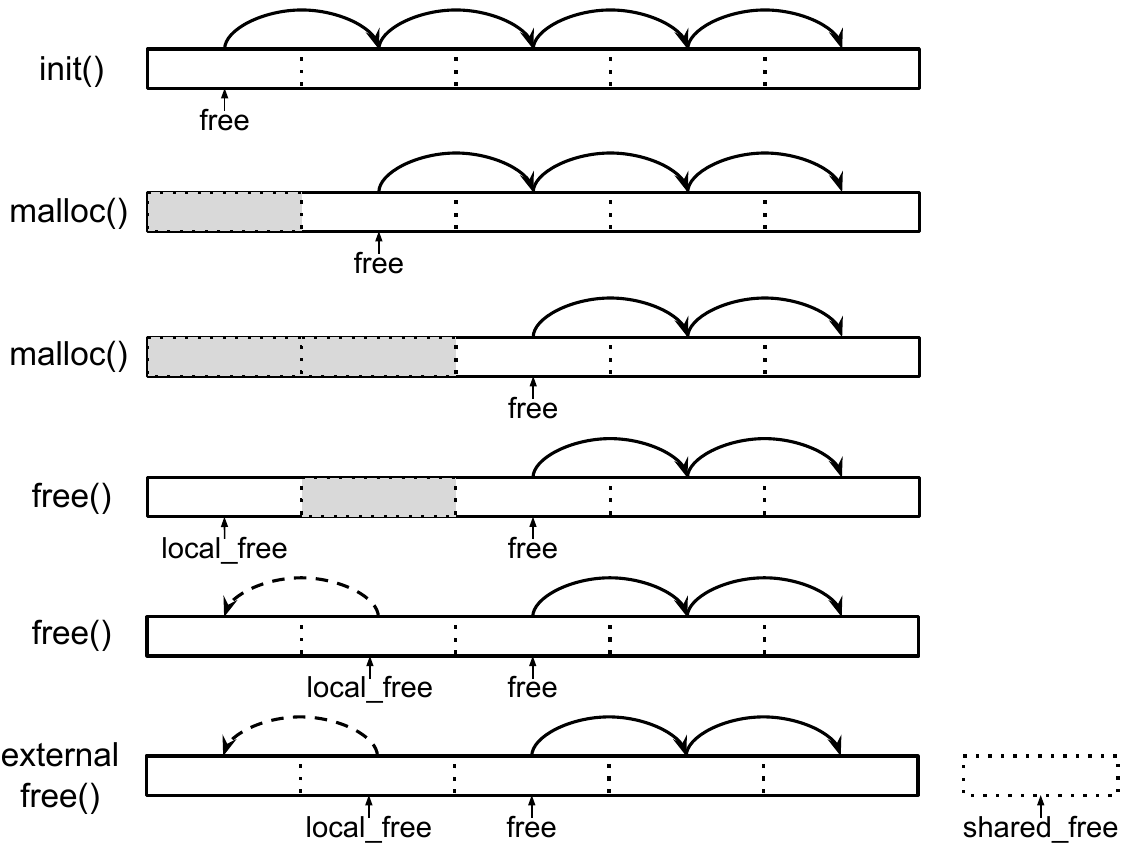}
	    \subcaption{Three free lists in mimalloc. }
	    \label{fig_free_list_1}
	\end{minipage} 
 	\begin{minipage}[t]{0.45\textwidth}
		\centering
		\includegraphics[width=\textwidth,trim = 5mm 4mm 92mm 0mm, clip=true, page=5]{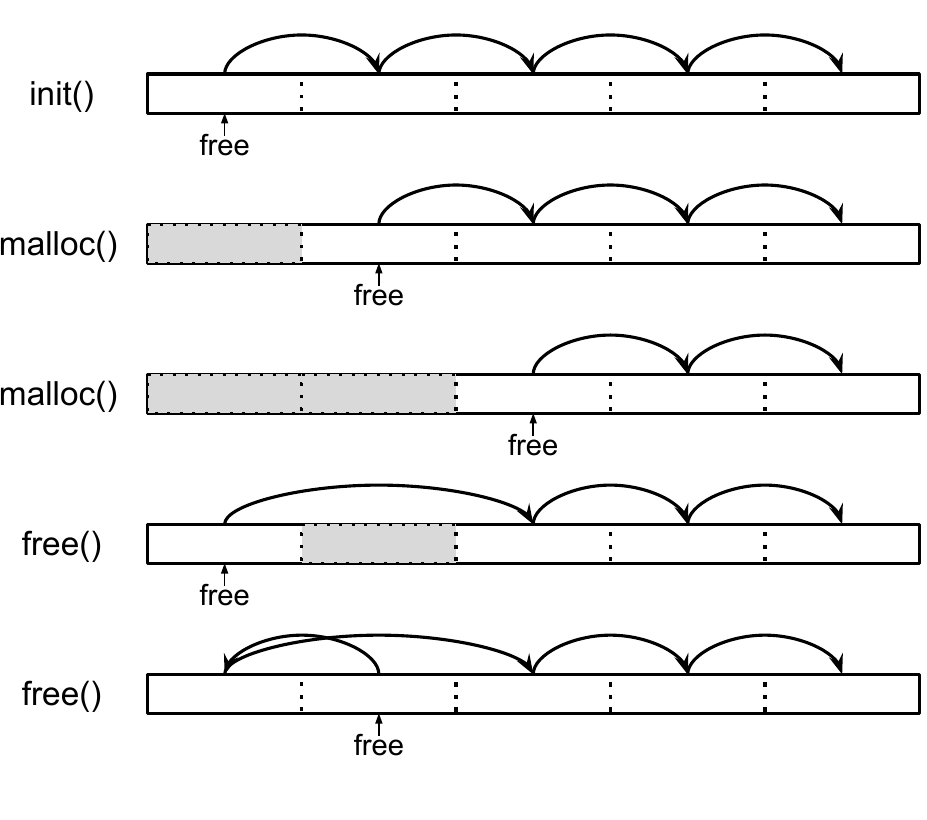}
		\subcaption{Single free list in \ExMalloc.}
	\label{fig_free_list_2}
	\end{minipage} 
 \caption{\ExMalloc uses a unified list of free, local free, and shared free lists, referred to as a single free list. }
 \label{fig_free_list}
\end{figure*}

\subsection{Single Free-Block List in Pages}
\label{section_no_local_free}
In addition to using a centralized heap to simplify overall metadata organization, \ExMalloc employs further techniques to simplify page-level metadata and enhance data access locality. 
Rather than maintaining multiple linked lists to track available free memory within each page (the lowest-level metadata structure in the hierarchy, as shown in Fig.~\ref{fig_heap_layout}), \ExMalloc maintains only a single free-block list\footnote{In this paper, we use the terms free-block list and free list interchangeably.} per page.\par

Modern memory allocators such as mimalloc maintain multiple free lists for each page, including one free list, one local free list, and one shared free list.
These structures are necessary to (a) balance memory usage across threads and prevent memory \textit{blowup}, and (b) handle cases of freeing memory allocated by a different thread.
As shown in Fig.~\ref{fig_free_list_1}, when a thread allocates a new page of memory, it populates all its memory blocks into the free list for immediate allocation. 
During allocation, the page assigned to the requesting thread first serves requests from its free list, avoiding costly locks on shared global metadata.
The local free list, in contrast, stores recently freed memory blocks that are not immediately available for allocation.
The shared free list stores external memory blocks allocated from another thread but freed in the current thread.
If the free list is exhausted, the allocator transfers all entries from the local free list and shared free list into the free list and attempts allocation again; if the local free list and shared free list are also empty, it then falls back to the global pool (either from the segment/heap or directly from the OS, depending on the allocator design). \par

Maintaining multiple free lists -- specifically, a free list, a local free list, and a shared free list -- helps reduce the frequency of synchronization operations. 
The allocator accesses the three lists independently and performs synchronization only when the free list becomes empty, thereby maintaining coherence while minimizing synchronization overhead.
However, this design can introduce additional sources of overhead.
Freed memory blocks may not be reused promptly, leading to reduced data locality.
Moreover, maintaining multiple free lists per page increases control complexity and the size of allocator metadata.
In contrast, \ExMalloc employs a single free list per page, which improves data locality and enables faster allocation and deallocation. \par

\noindent\textbf{Better data locality:}
Since \ExMalloc is designed for single-threaded workloads, maintaining three separate free lists is unnecessary in the absence of contention.
As shown in Fig.~\ref{fig_free_list_2}, a single (unified) free list is sufficient and more effective in preserving cache and TLB locality.
In contrast, multiple free list designs that defer reuse--by first pushing blocks to the local free list before returning them to the free list--can increase cache and TLB misses due to delayed reuse and page write-backs.
\par

\noindent\textbf{Faster allocation:}
Algorithms~\ref{alg_malloc_multi_free} and~\ref{alg_malloc_single_free} compare the design of multi-threaded allocators--where free lists, local free lists, and shared free lists are utilized--against \ExMalloc, which uses only a single free list. 
In addition to improving data locality, \ExMalloc simplifies control logic by checking just one list and eliminates synchronization overhead since no locks are required to handle contention across threads. 
As a result, \ExMalloc delivers faster allocations, reducing the so-called datacenter tax~\cite{kanev2017mallacc, hunter2021beyond, zhou2024characterizing}.

\noindent\textbf{Faster deallocation:}
Similar to the allocation process ($malloc()$), utilizing a single free list can also help achieve faster deallocation ($free()$). 
In multi-threaded allocators, freeing memory often requires additional bookkeeping to determine whether a page can be safely released, since blocks may be redistributed across shared free lists associated with pages in other threads.  
By contrast, \ExMalloc avoids this complexity entirely. 
With only one free list, it does not need to verify whether freed blocks belong to shared lists or whether other threads are still using a page.

\begin{figure}[htbp]
\centering
\begin{minipage}{0.9\linewidth}
\centering
\lstset{language=C}
\begin{lstlisting}[caption={Pseudo code of malloc() of using muliple free lists in multi-threaded allocators.}, label={alg_malloc_multi_free}]
void malloc(size_t size) {
    page = find the page based on block size;
    if (page->free != null) {
        void* p = page->free;
        page->free = page->free->next;
    }
    else if (page->local_free != null) {
        page->free = page->local_free;
        page->local_free = null;
        void* p = page->free;
    }
    else if (page->shared_free != null) {
        obtain thread lock();
        page->free = page->shared_free;
        page->shared_free = null;
        void* p = page->free;
        release thread lock();
    }
    else {
        page = new Page();
        void* p = page->free;
        page->free = page->free->next;
    }
    return p;
}
\end{lstlisting}
\end{minipage}
\end{figure}

\begin{figure}[htbp]
\centering
\begin{minipage}{0.9\linewidth}
\centering
\lstset{language=C}
\begin{lstlisting}[caption={Pseudo code of malloc() of using single free list in \ExMalloc.}, label={alg_malloc_single_free}]
void malloc(size_t size) {
    page = find the page based on block size;
    if (page->free != null) {
        void* p = page->free;
        page->free = page->free->next;
    }
    else {
        page = new Page();
        void* p = page->free;
        page->free = page->free->next;
    }
    return p;
}
\end{lstlisting}
\end{minipage}
\end{figure}

%% file: 03_03.tex
\subsection{Balance Memory Commitment and Reclamation}
\label{section_commit}

A well-designed memory allocator must balance allocation speed with overall memory consumption--a particularly challenging task in multi-threaded scenarios, where additional factors such as synchronization overhead and cross-thread memory balance come into play.
For single-threaded applications, however, this balance is more attainable. 
In \ExMalloc, we employ a carefully tuned strategy that balances memory commitment and reclamation to optimize both allocation performance and memory efficiency. \par

We set the segment size to 4 MiB to balance the allocator performance and memory fragmentation (an empirical analysis of this design choice is provided in \S~\ref{section_analysis}). 
For large pages (as described in \S~\ref{section_heap_layout}), each segment contains only one page, which bounds the worst-case internal fragmentation to 2 MiB. 
Increasing the segment size beyond this point may lead to higher fragmentation without providing significant performance benefits, particularly for applications that allocate memory infrequently but in large chunks -- hundreds of KiB or a few MiB per request (e.g., x264 in SPEC CPU2017, which encodes raw YUV video frames into the H.264/MPEG-4 AVC format).
These applications exhibit bursty allocation behavior, frequently creating and releasing multi-megabyte buffers, which makes their memory usage highly sensitive to allocator fragmentation.
A smaller segment size can reduce fragmentation; however, it may degrade performance -- particularly data locality -- since smaller segments often correspond to smaller page sizes. 
As a result, pages of the same block size may be distributed across multiple segments, leading to more frequent $mmap()$ system calls for segment allocation and increased overhead when traversing pages across segments to locate free blocks.
\par

Beyond selecting an appropriate segment size, \ExMalloc incorporates explicit mechanisms for segment memory commitment and reclamation between the allocator and the OS, as outlined in Algorithm~\ref{alg_commit}. 
For the initial segment allocated, \ExMalloc intentionally defers committing physical memory to the OS (achieved by committing memory at the per-page granularity), thereby accommodating applications with extremely small or short-lived memory footprints (e.g., exchange2 in SPEC CPU2017, which is a financial order-book matching engine). 
When a segment becomes free, it is first returned to the segment cache, which retains one cached segment per page type. This design avoids the overhead of repeated $mmap()$ and $munmap()$ system calls while enabling rapid segment reuse, ultimately improving both allocation latency and memory efficiency. \par

The commitment and reclamation mechanism in \ExMalloc is highly effective for single-threaded allocators.
In multi-threaded environments, allocators often maintain per-thread segment caches or deferred commitment segments to handle reclamation efficiently. 
However, these designs must balance between evenly distributed and highly skewed allocation behaviors across threads, resulting in complex design trade-offs.
State-of-the-art allocators, such as mimalloc, make these mechanisms user-configurable, which can pose challenges for application developers lacking deep knowledge of memory allocators.
In contrast, the single-threaded nature of \ExMalloc enables a unified, one-size-fits-all solution that is simpler, deterministic, and easier to tune for predictable performance and memory behavior. \par

\begin{figure}[htbp]
\centering
\begin{minipage}{0.9\linewidth}
\centering
\lstset{language=C}
\begin{lstlisting}[caption={Pseudo code of segment memory commiment and reclamation.}, label={alg_commit}]
void new_segment() {
    if (first segment) commit memory of each page on demand;
    else if (segment in cache) reclaim segment from cache;
    else commit memory to OS;
}

void free_segment() {
    if (cache not full) push segment to cache;
    else OS free segment;
}
\end{lstlisting}
\end{minipage}
\end{figure}

%% file: 03_04.tex
\subsection{Adopting Other Optimizations in Modern Multi-threaded Allocators}
\label{section_other}
Modern multi-threaded allocators employ several optimization techniques that are also beneficial in single-threaded contexts but are absent in legacy single-threaded allocators such as dlmalloc.
\ExMalloc incorporates these techniques to further enhance performance.

\noindent{\textbf{Aggregated metadata:}} Aggregated metadata in memory allocators refers to the consolidation of metadata structures -- such as free block information of each page -- into compact, contiguous memory regions~\cite{li2023nextgen, leijen2019mimalloc, reitz2024starmalloc}. This design is particularly well-suited for the single-threaded architecture of \ExMalloc. 
Aggregated metadata improves spatial locality by grouping related metadata entries with program user data, thereby reducing cache misses (the data in the allocated address is already cached when accessing metadata). 
Additionally, since \ExMalloc is specifically designed for single-threaded scenarios, both allocator metadata and user data remain physically localized to a single core. 
By retaining aggregated metadata, \ExMalloc eliminates the need for additional bits to store reference pointers for each free memory block, resulting in reduced memory consumption. \par

In contrast, segregated metadata may introduce additional overhead in performance-sensitive, single-threaded contexts, though useful in security-critical scenarios, where spatial separation of metadata and payload is required to mitigate metadata-based attacks~\cite{reitz2024starmalloc}. However, since the security-critical scenario is not the main focus of this paper, \ExMalloc uses aggregated metadata by default as it is performance-oriented. \par

\noindent{\textbf{Bitmaps:}} \ExMalloc employs compact metadata structures--such as bitmaps--to optimize common allocation and deallocation paths, enabling constant-time (O(1)) fast-path operations (e.g., indexing segments of different block sizes in Fig.~\ref{fig_heap_layout}). 
These structures allow the allocator to quickly identify available memory blocks without traversing linked lists or performing complex lookups. 
By leveraging such compact data representations, \ExMalloc achieves fast allocation while minimizing metadata overhead. \par

\noindent{\textbf{Fine-grained block-size classes:}} Using fine-grained block-size classes helps minimize internal fragmentation by more precisely aligning requested allocation sizes with actual block sizes, thereby reducing wasted memory within allocated blocks. 
This approach is particularly beneficial for small object allocations, where even slight mismatches between requested and allocated sizes can lead to significant cumulative overhead. 
For instance, dlmalloc classifies allocations smaller than 512 bytes into small bins with a 16-byte step size, meaning that a request for 33 bytes would be rounded up to 48 bytes. 
In contrast, \ExMalloc adopts a finer-grained 8-byte step size for its small size classes (e.g., the 8-byte and 16-byte block size in Fig.~\ref{fig_heap_layout}), enabling more precise matching of allocation sizes and thereby reducing internal fragmentation. \par

\noindent{\textbf{Inline functions:}} \ExMalloc leverages inline functions to help reduce the overhead associated with frequent allocation and deallocation calls by eliminating the cost of function dispatch. 
In traditional allocators, such as dlmalloc, each memory operation may involve a function call, which introduces additional instruction overhead and potential pipeline stalls--especially problematic in tight allocation loops where these calls occur repeatedly. 
By contrast, inlining replaces function calls with their actual implementation code during compilation, allowing the compiler to perform further optimizations such as constant folding, loop unrolling, or register allocation. \par

%% file: 04_using.tex
\subsection{Implementation of \ExMalloc}
We implemented \ExMalloc by stripping away multi-threaded metadata and synchronization logic to better serve the needs of single-threaded applications. 
The result is an extremely lightweight allocator comprising approximately 2,000 lines of C/C++ code. 
\ExMalloc can be compiled as either a static or shared library, similar to existing allocators such as jemalloc, tcmalloc, and mimalloc. 
It provides full support for standard C/C++ memory allocation and deallocation interfaces, including $malloc()$, $free()$, $calloc()$, and $realloc()$ in C, as well as $new$ and $delete$ in C++. \par

\begin{figure}[t]
    \centerline{\includegraphics[width=0.8\columnwidth, trim = 15mm 54mm 21mm 21mm, page=8, clip=true]{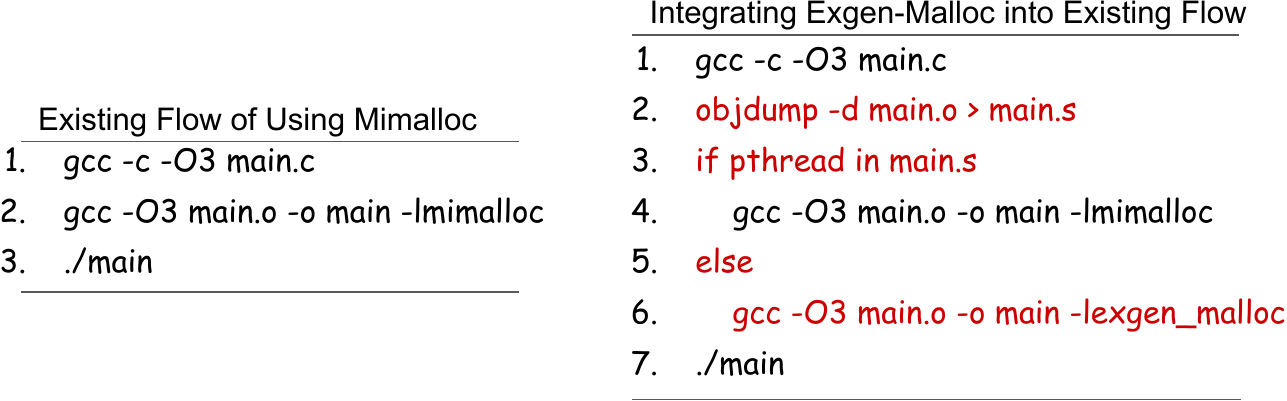}}
	\caption{Compilation flow with/without \ExMalloc.}
	\label{fig_compile}
\end{figure}

\subsection{Using \ExMalloc}
To ensure \ExMalloc is a practical solution, it must integrate seamlessly with existing compilation workflows.
This requires an understanding of how current systems incorporate custom memory allocators (e.g., mimalloc), typically by linking the allocator library and overriding the default memory allocation functions.
For example, mimalloc can be integrated either by linking its static library or by preloading its shared library as a drop-in replacement.
As illustrated in Fig.~\ref{fig_compile} (left), a common approach on POSIX-based systems using $gcc$ is to preload mimalloc in place of the default glibc allocator.
This workflow serves as a foundation for integrating \ExMalloc into the compilation process. \par

Integrating \ExMalloc into existing compilation workflows requires an additional step to determine whether a program is single- or multi-threaded. 
One way is to introduce an extra compilation pass.
As shown in Fig.~\ref{fig_compile} (right), after compiling \texttt{main.c} but before linking memory allocators to \texttt{main.o}, the process disassembles the object file to detect references to multi-threaded libraries such as POSIX threads (\texttt{pthread}).
This detection can be automated using shell scripts (or scripts in other languages) and incorporated into the build system (e.g., Makefile)\footnote{Before the integration of libpthread into the standard C library in glibc 2.34, it was even simpler to determine if a program like $main$ is multi-threaded by using $ldd$ $main$ to check if libpthread was linked to it.}. 
Importantly, developers can continue using standard memory allocation functions (e.g., \texttt{malloc()}/\texttt{free()}) without modifying their applications. 
This makes integrating \ExMalloc into existing workflows straightforward and efficient, enabling developers to benefit from its performance improvements without altering established development practices. \par

%% file: 05_00_evaluation.tex
We evaluate the effectiveness of \ExMalloc by addressing the following questions:

\begin{itemize}
\item How does the performance of \ExMalloc compare to other single-threaded and multi-threaded allocators? (\S~\ref{section_performance_results})
\item Can \ExMalloc reduce memory consumption compared to other single-threaded and multi-threaded allocators? (\S~\ref{section_memory_results})
\item Why did \ExMalloc achieve performance gains over other allocators? (\S~\ref{section_analysis})
\item How \ExMalloc determines its design parameters? (\S~\ref{section_sensitivity})
\end{itemize}

Some highlights of our results include:
\begin{itemize}
    \item \ExMalloc achieves $1.17\times$, $1.10\times$, $1.93\times$ speedups over dlmalloc on SPEC CPU2017, redis-benchmark, and mimalloc-bench, respectively.
    \item \ExMalloc achieves $6.2\%$, $0.1\%$, and $25.2\%$ memory savings over mimalloc on SPEC CPU2017, redis-benchmark, and mimalloc-bench, respectively.
\end{itemize}

\input{05_01}

\input{05_02}

\input{05_03}

\input{05_04}
\input{05_05}

%% file: 05_01.tex
\subsection{Evaluation Methodology}
\noindent{\textbf{Platforms:}}
To validate that using single-threaded memory allocators for single-threaded programs can lead to performance improvements, we evaluate \ExMalloc on two Intel Xeon systems -- System A with Skylake processors and System B with Sapphire Rapids processors (detailed machine configurations listed in Table~\ref{tab:spec_systems}). 

\begin{table}[t]
    \caption{Specifications of evaluation systems.}
    \label{tab:spec_systems}
    \centering
    \small
    \setlength{\tabcolsep}{3pt}
    \renewcommand{\arraystretch}{1.25}
    \begin{tabular}{|| >{\centering\arraybackslash}p{1.3cm} |
                       >{\centering\arraybackslash}p{5.8cm} |
                       >{\centering\arraybackslash}p{5.8cm} 
                       ||}
        \hline
        & {\bf System A (Skylake)} & {\bf System B (Sapphire Rapids)}   \\
        \hline
        \hline
        \textbf{Cores} & dual-socket 24$\times$ Intel Xeon Platinum 8160
        &
        dual-socket 56$\times$ Intel Xeon CPU Max 9480
        \\
        \hline
        \textbf{Caches} & 32KB L1D, 32KB L1I, 1MB L2, 33MB LLC &
         48KB L1D, 32KB L1I, 1MB L2, 112.5MB LLC \\
        \hline
        \textbf{DRAM} & 192GB DDR4 &
        128GB HBM2e \\
        \hline
        {\bf Kernel} & {\tt Linux 5.14}  &
        {\tt Linux 5.14}       \\
        \hline
        {\bf Compiler} & {\tt gcc 13.2.0}  &
        {\tt gcc 13.2.0}       \\
        \hline
    \end{tabular}
\end{table}

\noindent{\textbf{Baselines:}}
We compare \ExMalloc against the single-threaded allocator dlmalloc~\cite{lea1996memory}, the glibc (2.35) allocator (legacy multi-threaded allocator), and three state-of-the-art multi-threaded allocators -- jemalloc~\cite{evans2006scalable}, tcmalloc~\cite{tcmalloc}, and mimalloc~\cite{leijen2019mimalloc}.

\noindent{\textbf{Benchmarks:}}
We evaluate the efficacy of \ExMalloc on three widely used benchmark suites:
\begin{itemize}
    \item \textit{SPEC CPU2017}: we evaluate the efficacy of \ExMalloc on the SPEC CPU2017 (int rate) workloads~\cite{spec}, a standard benchmark suite broadly used for evaluating processors and compilers~\cite{panda2018wait, lee2018reconciling, mukherjee2020dataflow, livinskii2020random, poesia2020dynamic, weber2025relaxing} as well as memory allocators~\cite{aigner2015fast, kanev2017mallacc, lietar2019snmalloc}.
    We use the \textsl{ref} input sizes for all workloads, which are designed to reflect the intended production-level problem size for each benchmark~\cite{spec, zhou2024characterizing}.
    We run n concurrent copies -- each isolated on its own core, where n is the total number of cores in the system.
    \item \textit{redis-benchmark}: we evaluate the performance of Redis~\cite{redis}, which is a single-threaded in-memory database~\cite{sui2020flow2vec, spall2020build, banerjee2020sound, kallas2020diffstream}, using redis-benchmark~\cite{redis-benchmark}.
    The redis-benchmark is configured with 100,000 total requests, 10,000 clients, and a pipeline depth of 16, matching the configuration used in the mimalloc evaluation.
    \item \textit{mimalloc-bench}: we evaluate the mimalloc-bench suite~\cite{mimalloc-bench}\footnote{Benchmark suites for jemalloc and tcmalloc are not publicly available.}, a widely used collection that includes allocator stress tests and real-world applications, as also used in prior studies~\cite{leijen2019mimalloc, li2023nextgen, reitz2024starmalloc}.
We use the standard input sizes defined in mimalloc-bench, consistent with the original mimalloc evaluation, but restrict execution to a single thread or process.
\end{itemize}

%% file: 05_02.tex
\begin{figure}[t]
    \centerline{\includegraphics[width=\columnwidth, trim = 9mm 9mm 8mm 8mm, page=1, clip=true]{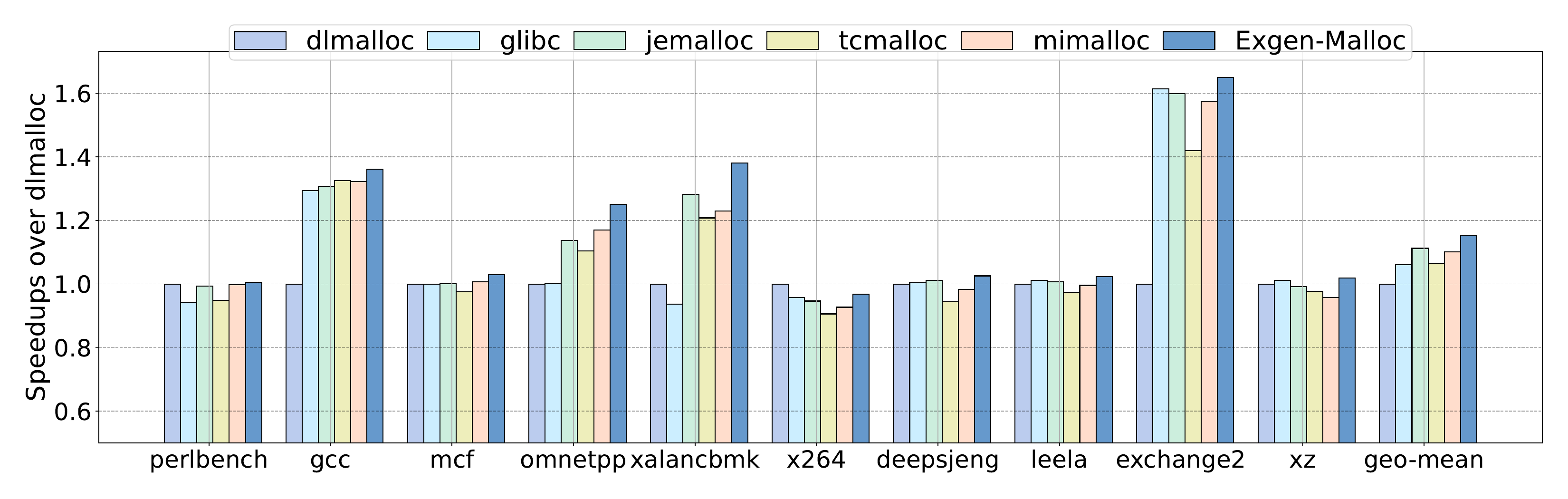}}
    \caption{Comparison of \ExMalloc against other allocators on SPEC CPU2017 (System A), showing speedups over dlmalloc -- higher values indicate better performance. 
    }
	\label{fig_results_spec_system_a}
\end{figure}

\begin{figure}[t]
    \centerline{\includegraphics[width=\columnwidth, trim = 9mm 9mm 8mm 8mm, page=1, clip=true]{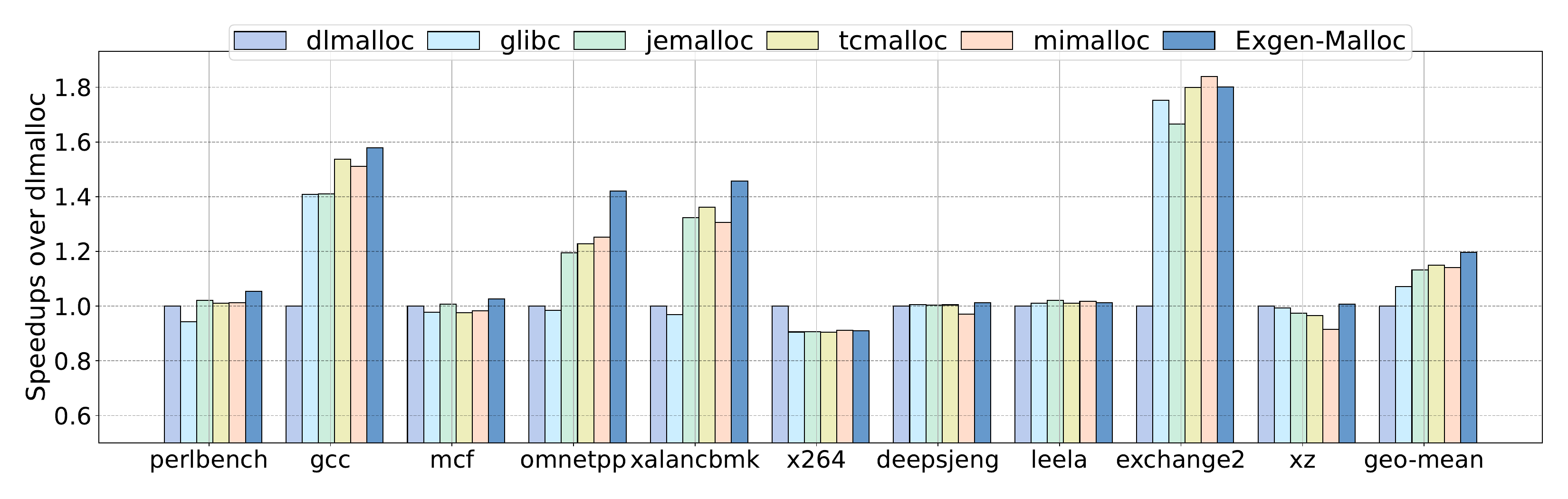}}
    \caption{Comparison of \ExMalloc against other allocators on SPEC CPU2017 (System B), showing speedups over dlmalloc -- higher values indicate better performance.  
    }
	\label{fig_results_spec_system_b}
\end{figure}

\subsection{Performance Results}
\label{section_performance_results}
Across all three benchmark suites, state-of-the-art multi-threaded allocators--jemalloc, tcmalloc, and mimalloc--consistently outperform the single-threaded allocator dlmalloc and the legacy multi-threaded glibc allocator. 
Their superior performance stems from advanced metadata structures that enhance data locality and reduce the overhead of frequent system calls.
By combining a simplified metadata design and specialized optimizations for single-threaded execution with selected techniques from modern multi-threaded allocators, \ExMalloc achieves the best overall performance among all allocators evaluated in this study.\par

\begin{figure}[t]
    \centerline{\includegraphics[width=0.95\columnwidth, trim = 9mm 9mm 8mm 8mm, page=1, clip=true]{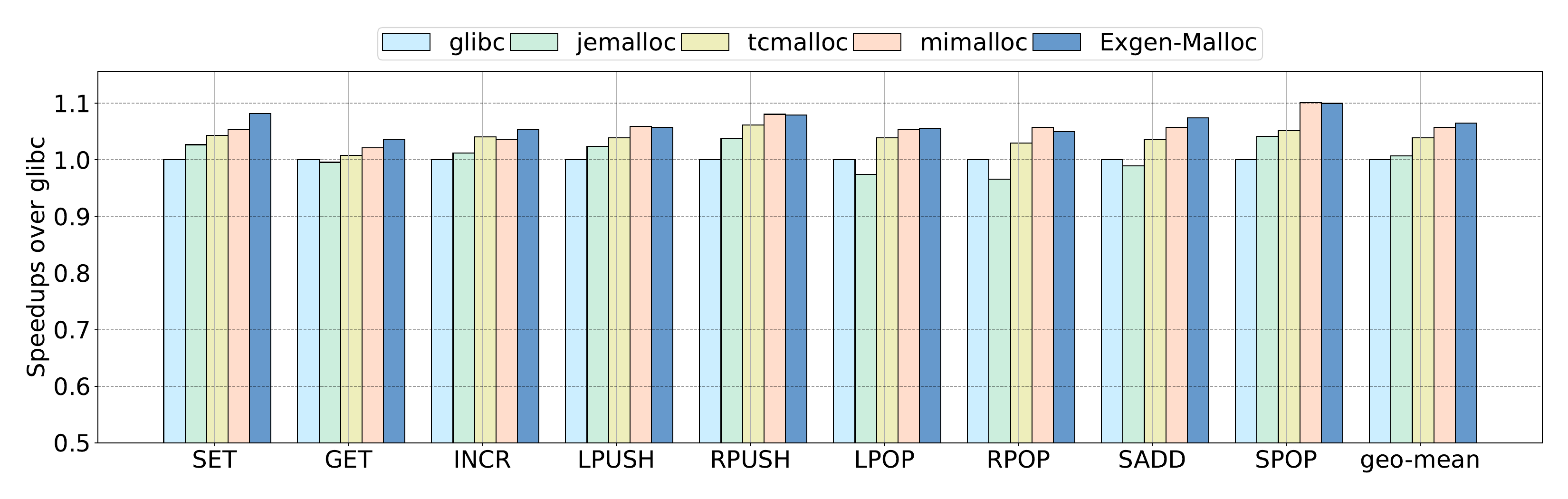}}
    \caption{Performance comparison of \ExMalloc and other allocators on redis-benchmark (speedups over glibc allocator because dlmalloc crashed on redis-benchmark). Higher values indicate better performance on redis-benchmark running on System A.
    }
	\label{fig_results_redis_system_a}
\end{figure}

\begin{figure}[t]
    \centerline{\includegraphics[width=0.95\columnwidth, trim = 9mm 9mm 8mm 8mm, page=1, clip=true]{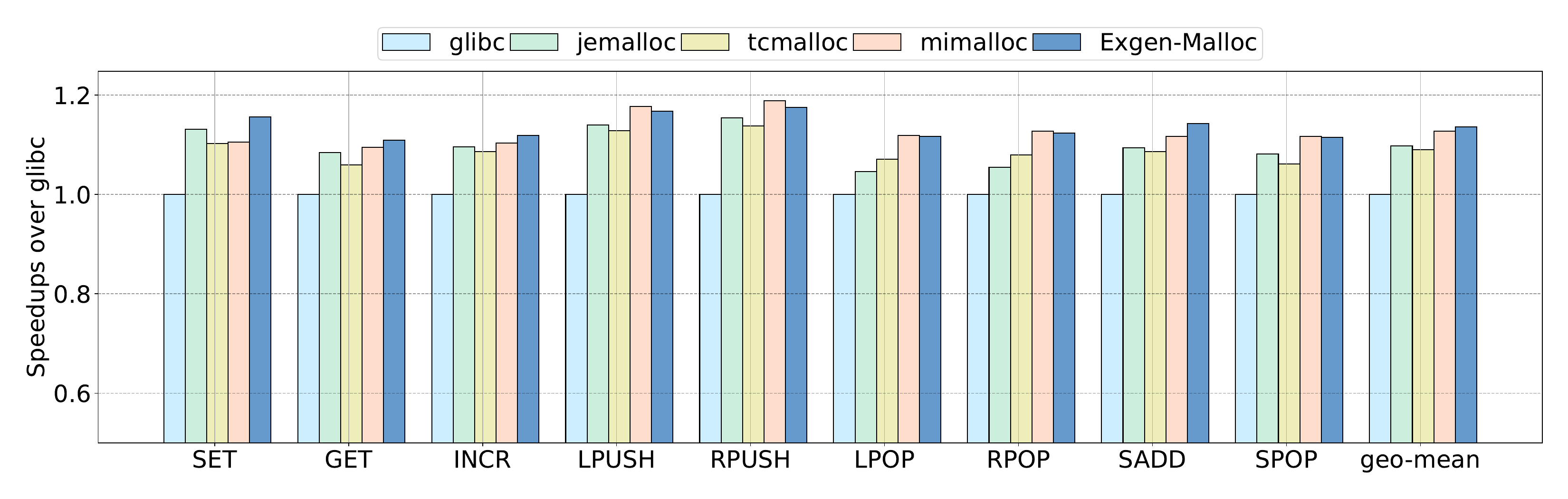}}
    \caption{Performance comparison of \ExMalloc and other allocators on redis-benchmark (speedups over glibc allocator because dlmalloc crashed on redis-benchmark). Higher values indicate better performance on redis-benchmark running on System B.
    }
	\label{fig_results_redis_system_b}
\end{figure}

\subsubsection{Results on SPEC CPU2017}
To evaluate \ExMalloc, we compare it against dlmalloc, the glibc allocator, jemalloc, tcmalloc, and mimalloc using SPEC CPU2017.
Fig.~\ref{fig_results_spec_system_a} shows the results on System A, while Fig.~\ref{fig_results_spec_system_b} shows those on System B.
Although jemalloc~\cite{evans2006scalable}\footnote{We compile jemalloc using the lazy\_lock flag, enabling it to operate in single-threaded mode without incurring mutex locking overhead. 
While this reduces synchronization costs, the underlying metadata structures remain as complex as in the multi-threaded configuration. 
}, tcmalloc~\cite{leijen2019mimalloc}, and mimalloc~\cite{leijen2019mimalloc} are multi-threaded, they outperform the single-threaded dlmalloc~\cite{lea1996memory} due to applying advanced design principles in modern allocators (\S~\ref{section:design}). 
By adopting modern allocator design principles while preserving a single-threaded architecture, \ExMalloc consistently outperforms existing allocators, achieving a geometric mean speedup of $1.15 \times$ over dlmalloc, $1.09 \times$ over glibc allocator, $1.04 \times$ over jemalloc, $1.08 \times$ over tcmalloc, and $1.05 \times$ over mimalloc on System A. 
On System B, where the case is of more advanced CPU architectures, \ExMalloc also outperforms existing allocators, achieving a geometric mean speedup of $1.20 \times$ over dlmalloc, $1.12 \times$ over glibc allocator, $1.06 \times$ over jemalloc, $1.04 \times$ over tcmalloc, and $1.05 \times$ over mimalloc.
Across all SPEC CPU2017 workloads, \ExMalloc shows the greatest performance improvements on \textsl{gcc}, \textsl{omnetpp}, and \textsl{xalancbmk}. 
These gains stem from its simplified metadata structure and single free list design (described in \S~\ref{section_no_local_free}), which enhances data locality (further analysis is provided in \S~\ref{section_analysis}).
\par

\subsubsection{Results on redis-benchmark}
\label{redis_result}
We also evaluate the effectiveness of \ExMalloc by comparing it to glibc allocator, jemalloc, tcmalloc, and mimalloc using redis-benchmark~\cite{redis-benchmark}, a benchmark suite to evaluate the performance of the single-threaded in-memory database Redis~\cite{redis}.
We did not include dlmalloc results since it crashed when running redis-benchmark.
The performance of \ExMalloc on System A and System B is shown in Fig.~\ref{fig_results_redis_system_a} and Fig.~\ref{fig_results_redis_system_b}, respectively. 
On System A, \ExMalloc achieves a geometric mean speedup of $1.07 \times$ over glibc allocator, $1.06 \times$ over jemalloc, $1.03 \times$ over tcmalloc, and $1.01 \times$ over mimalloc.
On System B, \ExMalloc achieves a geometric mean speedup $1.14 \times$ over glibc allocator, $1.04 \times$ over jemalloc, $1.04 \times$ over tcmalloc, and $1.01 \times$ over mimalloc.
\ExMalloc shows relatively smaller performance gains on redis-benchmark compared to SPEC CPU2017. 
This is because, in redis-benchmark, all query requests have identical sizes, resulting in a highly regular and predictable allocation pattern that can already be efficiently handled by modern allocators such as mimalloc. 
In contrast, real-world Redis workloads typically involve queries of varying lengths, where \ExMalloc would likely yield greater performance improvements~\cite{wu2020ac, wang2023catalyst}.

\begin{figure}[t]
    \centerline{\includegraphics[width=0.95\columnwidth, trim = 9mm 9mm 8mm 8mm, page=1, clip=true]{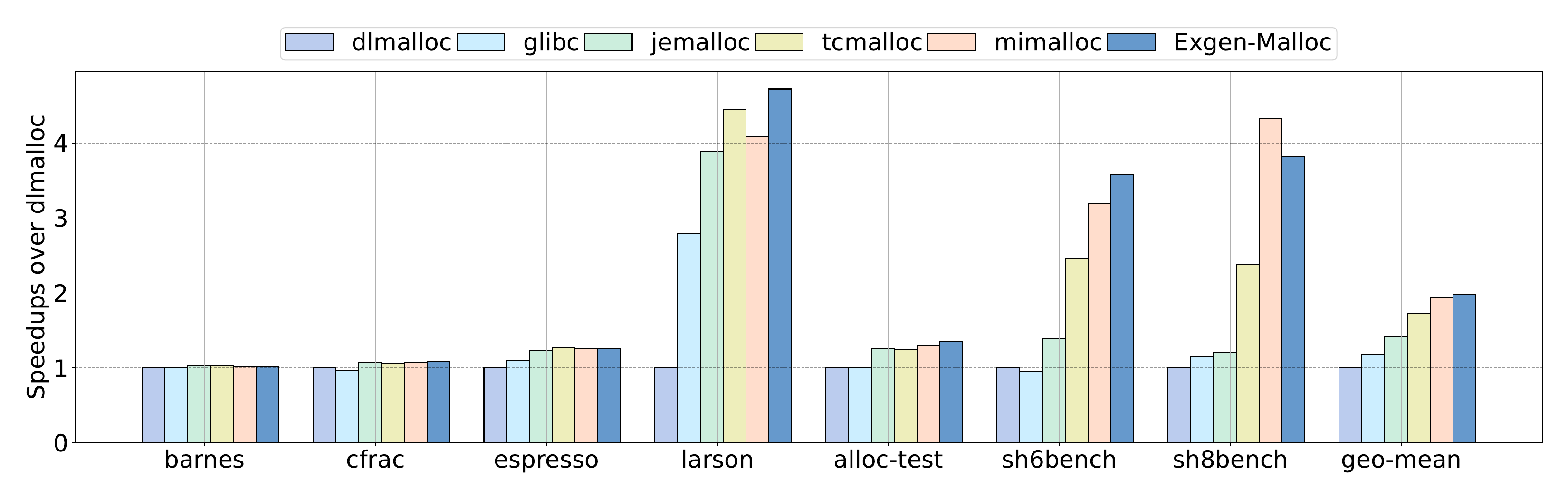}}
    \caption{Performance comparison of \ExMalloc and other allocators on mimalloc-bench (speedup over dlmalloc). Higher values indicate better performance on mimalloc-bench running on System A.
    }
	\label{fig_results_mimalloc_system_a}
\end{figure}

\begin{figure}[t]
    \centerline{\includegraphics[width=0.95\columnwidth, trim = 9mm 9mm 8mm 8mm, page=1, clip=true]{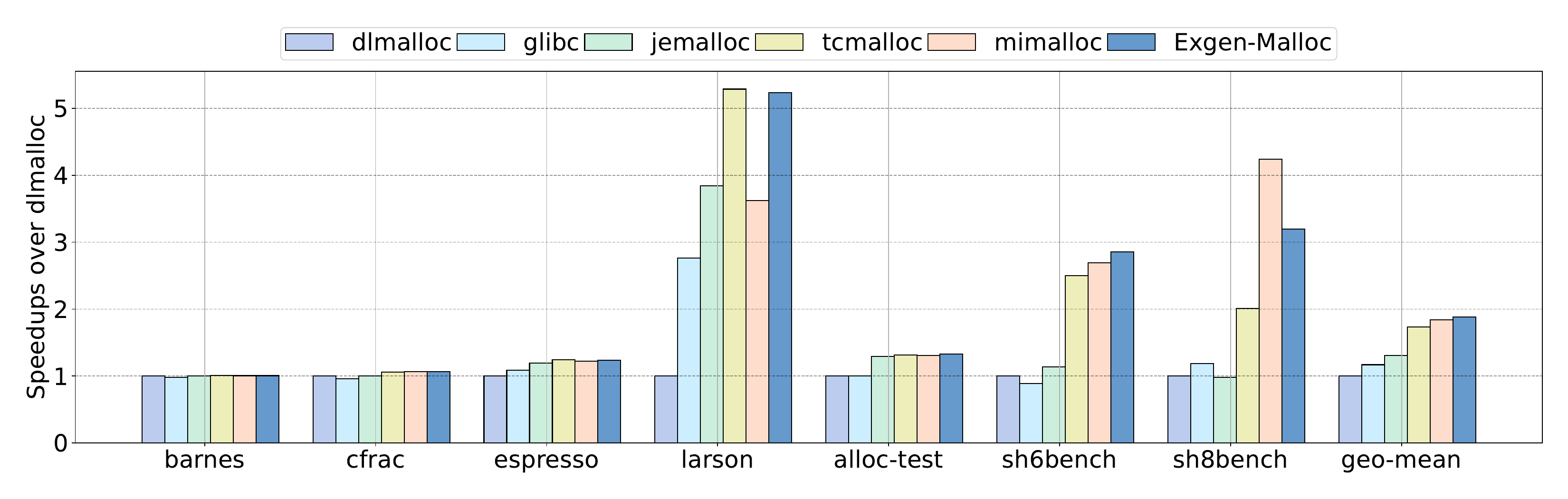}}
    \caption{Performance comparison of \ExMalloc and other allocators on mimalloc-bench  (speedup over dlmalloc). Higher values indicate better performance on mimalloc-bench running on System B.
    }
	\label{fig_results_mimalloc_system_b}
\end{figure}

\subsubsection{Results on mimalloc-bench}
We further evaluate the performance of \ExMalloc using mimalloc-bench\footnote{We exclude workloads in mimalloc-bench that specifically target multi-threaded behaviors--for example, cache-scratch and cache-thrash, which test cache line false sharing across threads and cores, and xmalloc-test, which evaluates freeing memory allocated by another thread.}, with results on System A and System B presented in Fig.~\ref{fig_results_mimalloc_system_a} and Fig.~\ref{fig_results_mimalloc_system_b}, respectively.
On System A, \ExMalloc achieves a geometric mean speedup of $1.98 \times$ over dlmalloc, $1.67 \times$ over glibc allocator, $1.41 \times$ over jemalloc, $1.15 \times$ over tcmalloc, and $1.03 \times$ over mimalloc.
On System B, \ExMalloc achieves a geometric mean speedup of $1.88 \times$ over dlmalloc, $1.61 \times$ over glibc allocator, $1.44 \times$ over jemalloc, $1.09 \times$ over tcmalloc, and $1.02 \times$ over mimalloc.
On real-world applications (within mimalloc-bench) such as \textsl{barnes}, \textsl{cfrac}, \textsl{espresso}, and \textsl{larson}, \ExMalloc consistently outperforms other allocators.
\ExMalloc may achieve slightly worse performance than mimalloc on microbenchmarks (within mimalloc-bench) like \textsl{sh6bench} and \textsl{sh8bench}, which are highly allocation-intensive. 
However, these patterns are less representative compared to real-world applications, as well as those in the SPEC CPU2017 standard benchmark suite.

%% file: 05_03.tex
\subsection{Memory Consumption Results}
\label{section_memory_results}
In addition to performance improvements, \ExMalloc also reduces memory consumption compared to state-of-the-art multi-threaded allocators--an equally important metric in evaluating the allocator efficiency.
In this section, we compared the memory consumption of \ExMalloc with other allocators on SPEC CPU2017, redis-benchmark, and mimalloc-bench as well. 
Since both systems share identical OS and compiler configurations, their memory consumption results are the same; thus, we report only those from System A. 
Across all benchmark suites, \ExMalloc delivers substantial performance improvements over dlmalloc and the glibc allocator, along with noticeable gains compared to state-of-the-art multi-threaded allocators--jemalloc, tcmalloc, and mimalloc.
In terms of memory utilization, \ExMalloc achieves significant savings relative to jemalloc, tcmalloc, and mimalloc, while incurring only slightly higher memory consumption than dlmalloc and the glibc allocator.
\par

\begin{figure}[t]
    \centerline{\includegraphics[width=\columnwidth, trim = 9mm 7mm 8mm 5mm, page=1, clip=true]{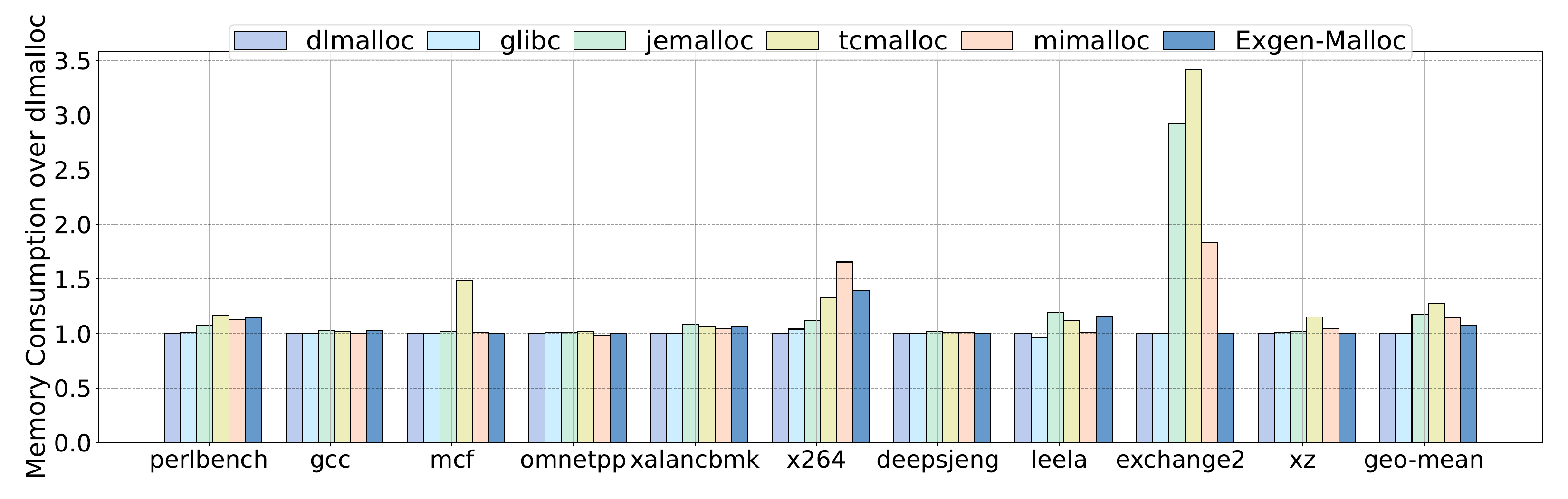}}
    \caption{Memory usage comparison between \ExMalloc and other allocators (normalized to dlmalloc; lower is better) on SPEC CPU2017 workloads.  
    }
	\label{fig_memprof_spec}
\end{figure}

\subsubsection{Results on SPEC CPU2017}
Fig.~\ref{fig_memprof_spec} presents the memory consumption results on SPEC CPU2017.
\ExMalloc achieves notable memory savings compared to state-of-the-art allocators. Specifically, it reduces memory usage by a geometric mean of $8.5\%$, $15.2\%$, and $6.2\%$ relative to jemalloc, tcmalloc, and mimalloc, respectively.
Compared to dlmalloc and the glibc allocator, \ExMalloc incurs slightly higher memory usage--$7.4\%$ and $7.0\%$ on average--primarily due to the x264 workload.
The x264 benchmark performs relatively fewer memory allocation requests than other workloads, but many of them are in the hundreds-of-kilobytes to a few-megabytes range. 
These medium-to-large allocations cause fragmentation within each 4 MiB segment (as defined in \S~\ref{section_heap_layout}).
Reducing the segment size could alleviate fragmentation but would likely degrade performance for other applications, as most workloads predominantly allocate small objects~\cite{wang2023memento, zhou2024characterizing, kanev2017mallacc}.
This trade-off explains why dlmalloc performs particularly well on x264 (Fig.~\ref{fig_results_spec_system_a} and Fig.~\ref{fig_results_spec_system_b}), but not on other workloads frequently allocating small objects.
\par

\subsubsection{Results on Redis-benchmark}
On the redis-benchmark, \ExMalloc also achieves good memory savings, as shown in Fig.~\ref{fig_memprof_spec}.
It reduces memory usage by a geometric mean of $0.1\%$, $23.1\%$, $31.7\%$, and $0.01\%$ compared to the glibc allocator, jemalloc, tcmalloc, and mimalloc, respectively (dlmalloc crashed during redis-benchmark).
These improvements stem from the fact that redis-benchmark features highly uniform allocation sizes (as discussed in \S~\ref{redis_result}), which minimizes fragmentation -- particularly benefiting allocators with aggregated metadata layouts such as mimalloc and \ExMalloc.

\begin{figure}[t]
    \centerline{\includegraphics[width=0.95\columnwidth, trim = 9mm 8mm 8mm 6mm, page=1, clip=true]{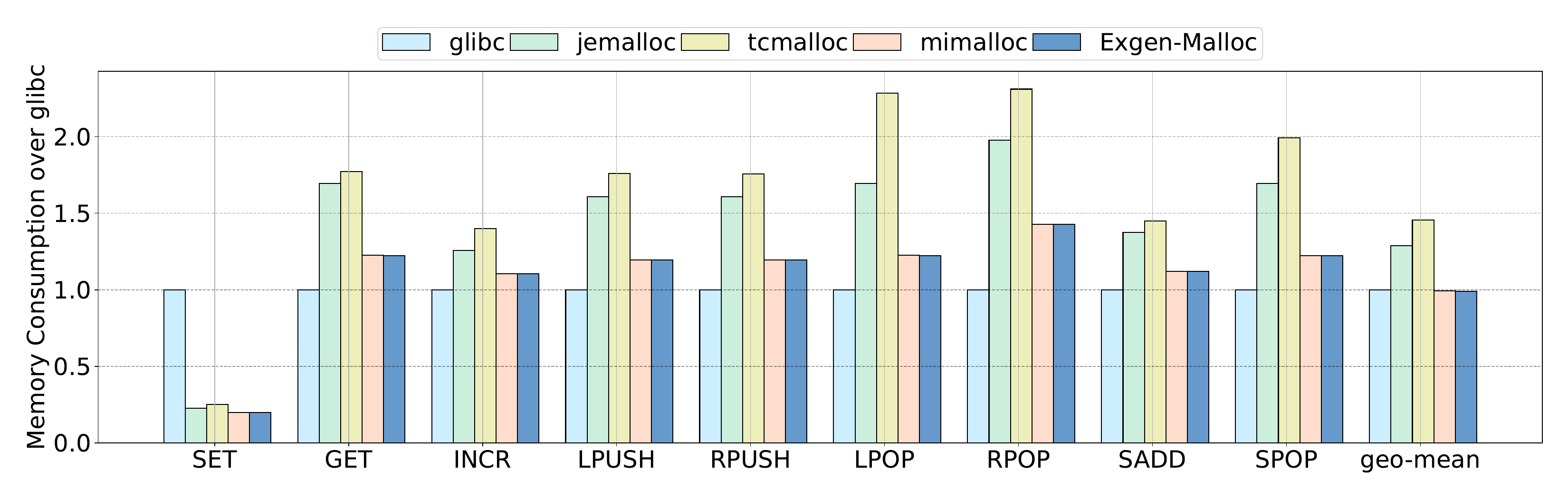}}
    \caption{Memory usage comparison between \ExMalloc and other allocators, normalized to glibc (lower is better), on redis-benchmark. dlmalloc crashed during the experiment.
    }
	\label{fig_memprof_redis}
\end{figure}

\begin{figure}[t]
    \centerline{\includegraphics[width=0.95\columnwidth, trim = 9mm 7mm 8mm 5mm, page=1, clip=true]{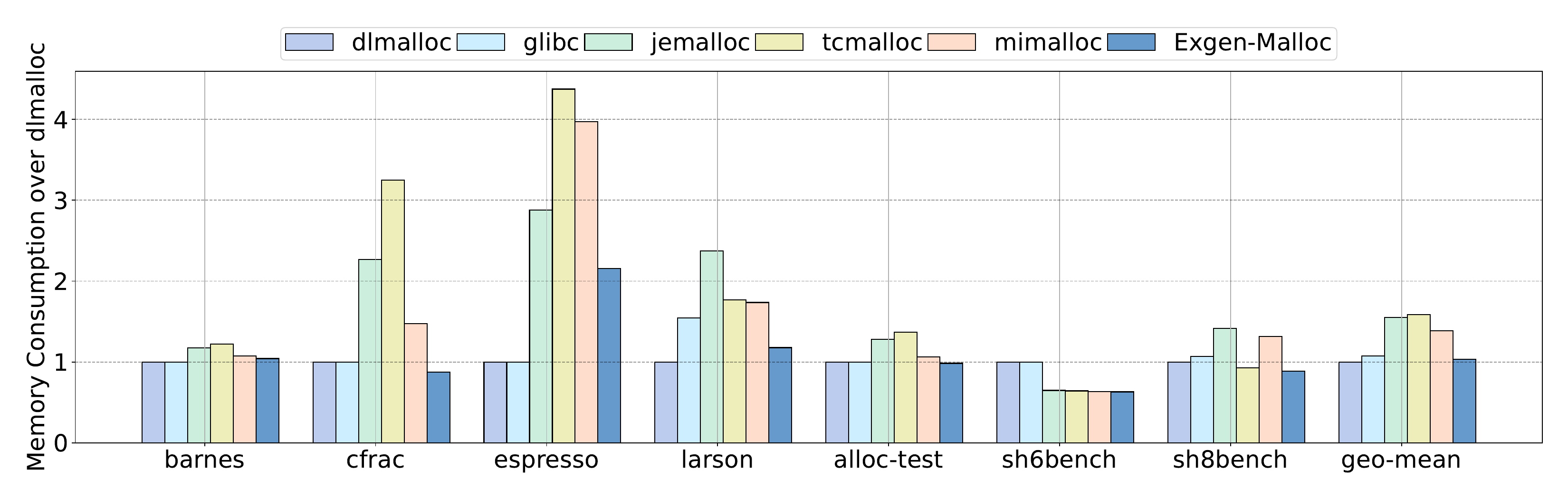}}
    \caption{Memory usage comparison between \ExMalloc and other allocators, normalized to glibc (lower is better), on mimalloc-bench. 
    }
	\label{fig_memprof_mimalloc}
\end{figure}

\subsubsection{Results on Mimalloc-bench}
We further evaluate the memory consumption of \ExMalloc on mimalloc-bench.
\ExMalloc exhibits slightly higher memory usage ($3.5\%$) than dlmalloc but achieves substantial memory savings compared to other allocators.
Specifically, it reduces memory usage by a geometric mean of $3.7\%$, $33.2\%$, $34.7\%$, and $25.2\%$ relative to the glibc allocator, jemalloc, tcmalloc, and mimalloc, respectively.
These gains arise because, in microbenchmarks -- particularly allocation-intensive ones like \textsl{sh6bench} and \textsl{sh8bench} -- the simplified metadata design of \ExMalloc eliminates much of the overhead associated with the complex multi-threaded mechanisms used in modern allocators, thereby improving memory efficiency. \par

%% file: 05_04.tex
\subsection{Performance Gain Analysis via Hardware Counters}
\label{section_analysis}
For allocation-intensive workloads with diverse object sizes such as \textsl{gcc}, \textsl{omnetpp}, and \textsl{xalancbmk}~\cite{akritidis2010cling, li2023nextgen} in SPEC CPU2017, \ExMalloc delivers substantial performance improvements over modern multi-threaded allocators ($1.08\times$, $1.14\times$, and $1.12\times$ speedups over jemalloc, tcmalloc, and mimalloc of \textsl{xalancbmk} on System A).
To further understand the source of the performance gains by using \ExMalloc, we profiled the three workloads under different memory allocators using hardware performance counters. \par

\begin{figure}[t]
	\begin{minipage}{0.49\textwidth}
		\centering
	\includegraphics[width=\textwidth, trim = 14mm 10mm 12mm 9mm, clip=true, page=1]{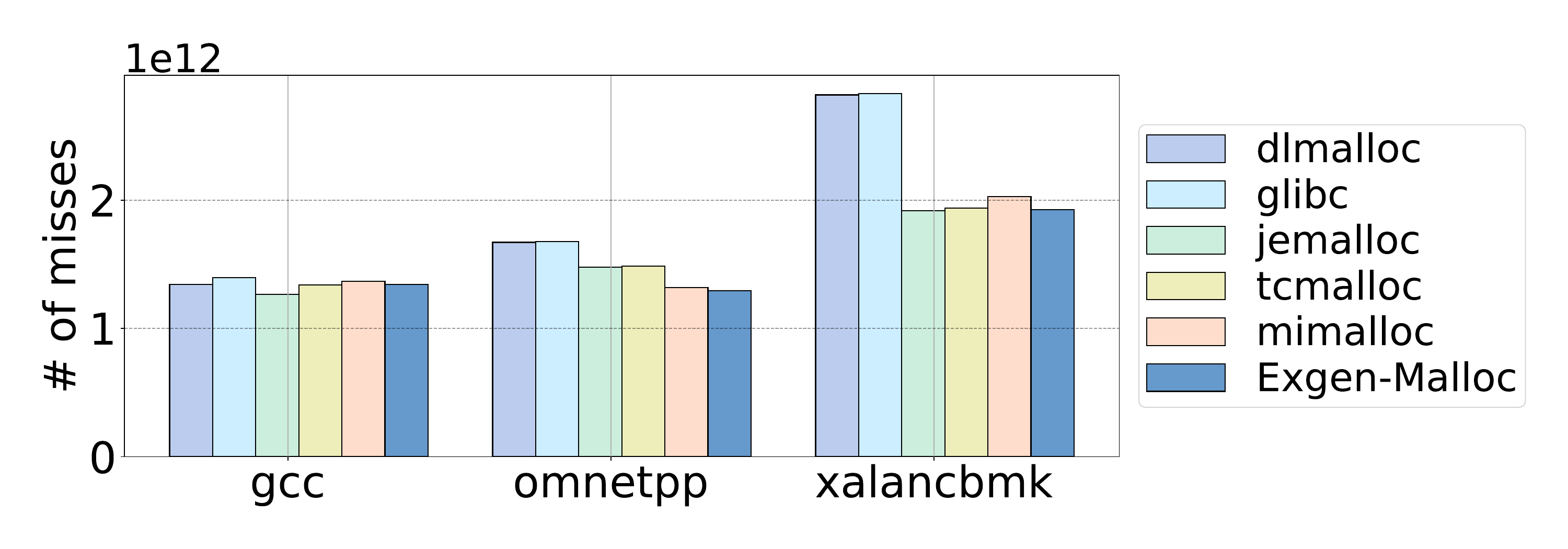}
	    \subcaption{L1 data cache misses (lower is better). }
	    \label{fig_breakdown_1}
	\end{minipage} 
 	\begin{minipage}{0.49\textwidth}
		\centering
		\includegraphics[width=\textwidth, trim = 14mm 10mm 12mm 9mm, clip=true, page=1]{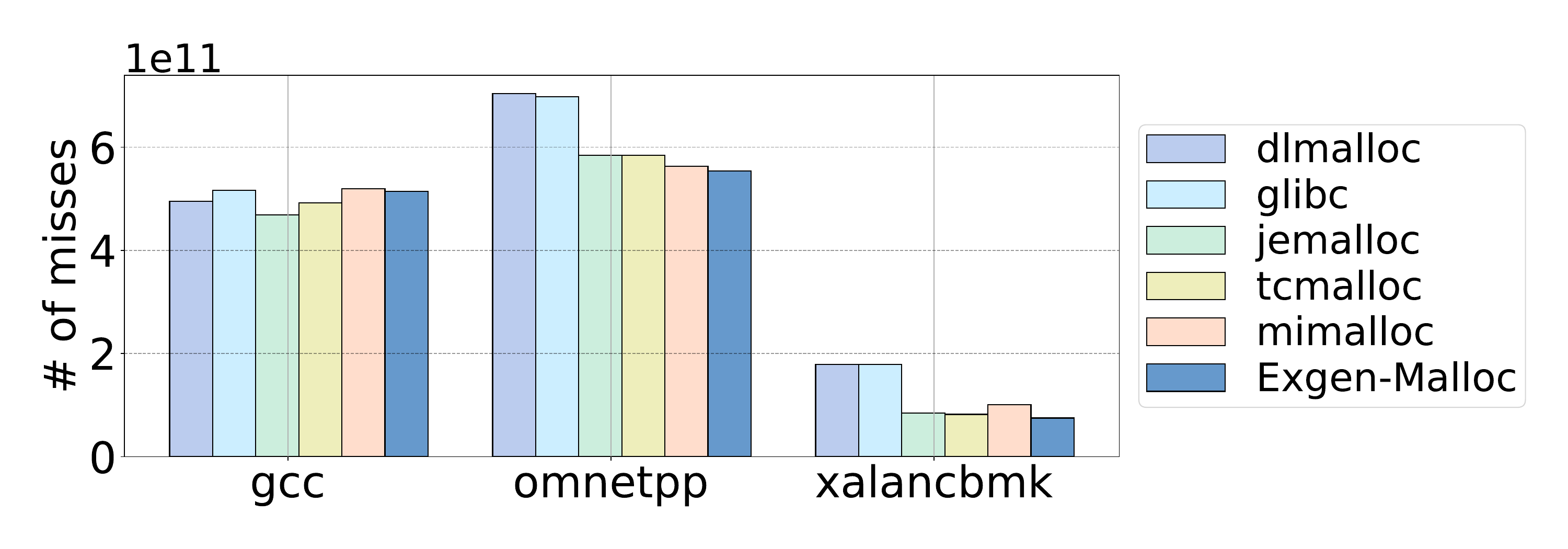}
		\subcaption{L2 cache misses (lower is better).}
	\label{fig_breakdown_2}
	\end{minipage} 
    \begin{minipage}{0.49\textwidth}
		\centering
		\includegraphics[width=\textwidth, trim = 14mm 10mm 12mm 9mm, clip=true, page=1]{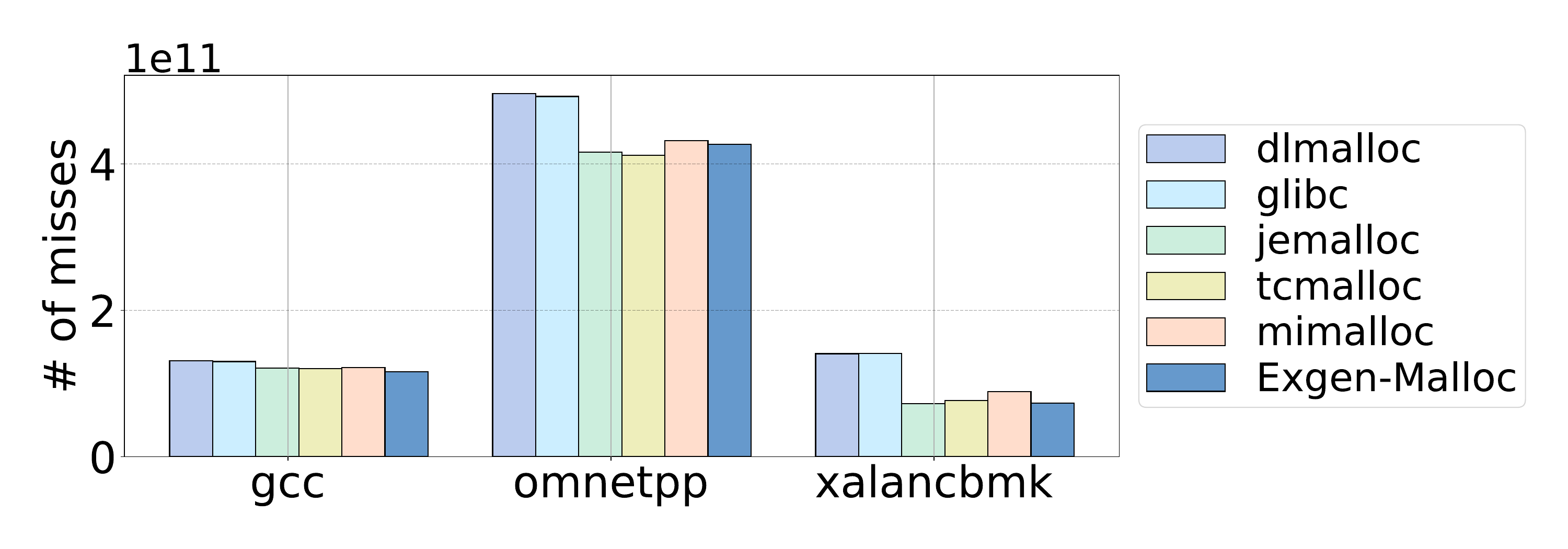}
		\subcaption{Last-level cache misses (lower is better).}
	\label{fig_breakdown_3}
	\end{minipage} 
    \begin{minipage}{0.49\textwidth}
		\centering
		\includegraphics[width=\textwidth,trim = 14mm 10mm 12mm 9mm, clip=true, page=1]{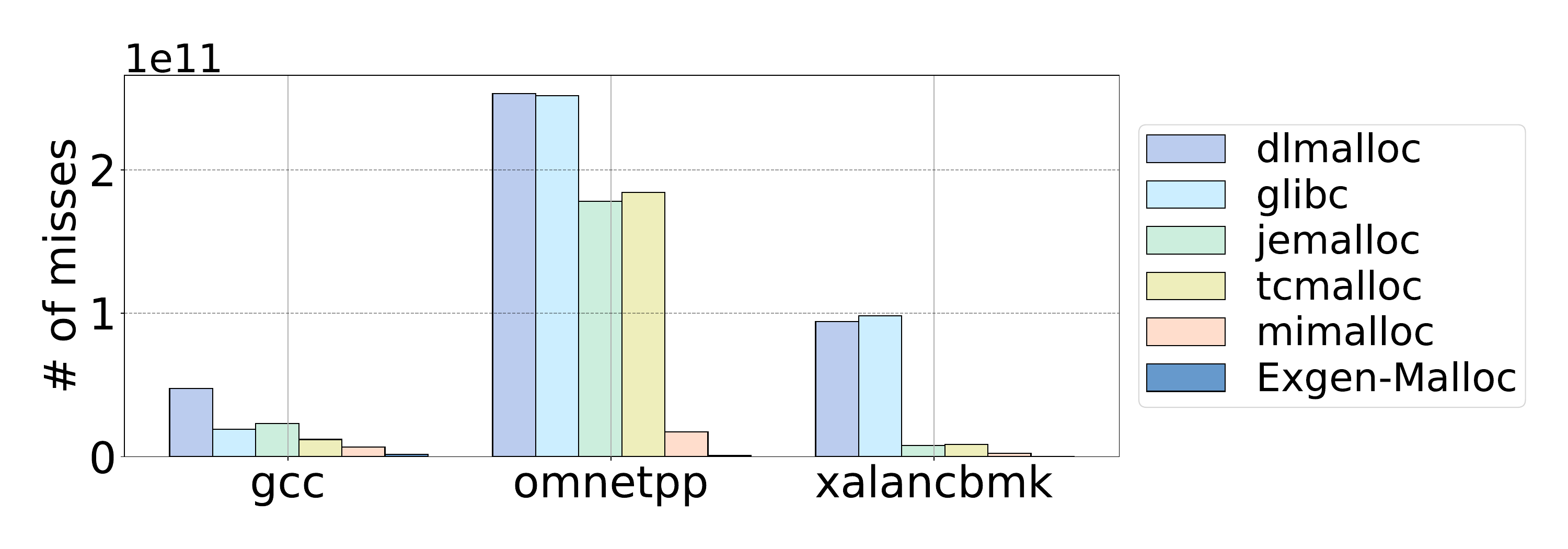}
		\subcaption{Data TLB misses (lower is better).}
	\label{fig_breakdown_4}
	\end{minipage} 
 \caption{Analyzing the factors leading to the performance gains in \ExMalloc by using hardware performance counters. \ExMalloc achieves better performance than other allocators by reducing L1 data cache, L2 cache, Last-level cache, and data TLB misses. }
 \label{fig_breakdown}
\end{figure}

We use L1 data cache misses, L2 cache misses, last-level cache misses, and data TLB misses as the main performance metrics, as these events exhibit strong correlation with program performance when the memory allocator is the primary bottleneck~\cite{maas2021adaptive, hunter2021beyond, li2023nextgen, zhou2024characterizing, zhou2023memperf}.
We omit instruction cache and instruction TLB related performance counters in this study, since these metrics are less correlated with data access behaviors.  \par

The reduction in cache misses indicates that \ExMalloc improves performance by minimizing allocator metadata and simplifying control logic. 
By employing a single free list that preferentially reuses recently freed objects, \ExMalloc enhances data locality. 
In addition to improving cache efficiency, this design also reduces TLB misses, as allocations are more likely to occur within pages that have been recently accessed rather than being deferred to per-thread free lists for future reuse.
As shown in Fig.~\ref{fig_breakdown_1}, \ExMalloc reduces L1 data cache misses by $18.0\%$, $19.5\%$, $1.9\%$, $4.4\%$, and $2.9\%$ compared to dlmalloc, the glibc allocator, jemalloc, tcmalloc, and mimalloc, respectively.
Similarly, Fig.~\ref{fig_breakdown_2} shows that \ExMalloc lowers L2 cache misses by $25.2\%$, $26.4\%$, $2.2\%$, $3.1\%$, and $9.4\%$ against the same baselines.
As shown in Fig.~\ref{fig_breakdown_3}, it further reduces last-level cache misses by $24.4\%$, $23.9\%$, $1.59\%$, and $7.8\%$ relative to dlmalloc, the glibc allocator, tcmalloc, and mimalloc, while performing only $0.04\%$ worse than jemalloc.
Finally, Fig.~\ref{fig_breakdown_4} illustrates that \ExMalloc achieves the largest reduction in data TLB misses—$98.6\%$, $96.8\%$, $96.5\%$, $94.3\%$, and $87.1\%$ lower than dlmalloc, the glibc allocator, jemalloc, tcmalloc, and mimalloc, respectively.
These reductions in data cache and TLB misses achieved by \ExMalloc confirm that the excessive allocator metadata and synchronization logic in modern multi-threaded allocators can degrade overall program performance.
\par

%% file: 05_05.tex
\subsection{Sensitivity Analysis of \ExMalloc Design Choices}
\label{section_sensitivity}
In this section, we perform a sensitivity analysis of our design choices in \ExMalloc, including segment size, page size, and the memory commitment and reclamation mechanisms (introduced in \S~\ref{section_commit}).
Although these parameters were initially chosen based on intuition, we provide empirical evidence to justify and validate our design decisions\footnote{We determine the design choices of \ExMalloc based on the geometric mean results of the SPEC CPU2017 benchmark suite, a widely used standard suite that captures diverse application behaviors. However, for application-specific optimizations, these choices can be adapted to better match the allocation patterns of individual workloads.}.

\begin{figure}[t]
	\begin{minipage}{\textwidth}
		\centerline{\includegraphics[width=0.95\columnwidth, trim = 9mm 9mm 8mm 8mm, page=1, clip=true]{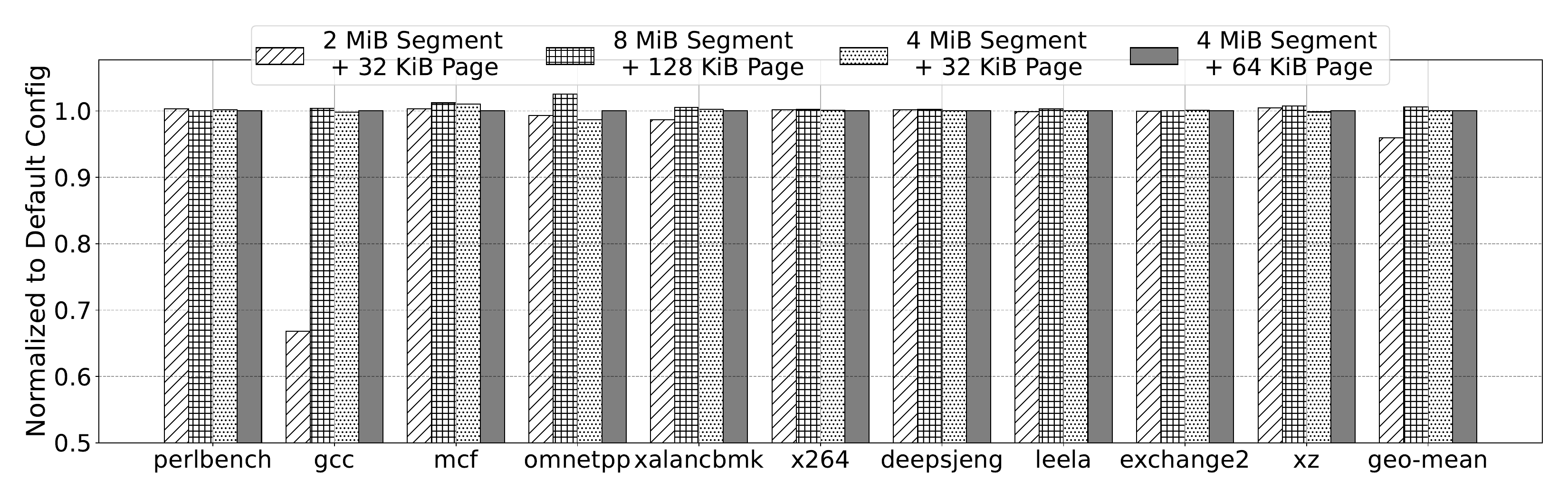}}
	    \subcaption{Performance on SPEC CPU2017 (larger is better). }
	    \label{fig_sensitivity_page_perf}
	\end{minipage} 
 	\begin{minipage}{\textwidth}
		\centerline{\includegraphics[width=0.95\columnwidth, trim = 9mm 9mm 8mm 8mm, page=1, clip=true]{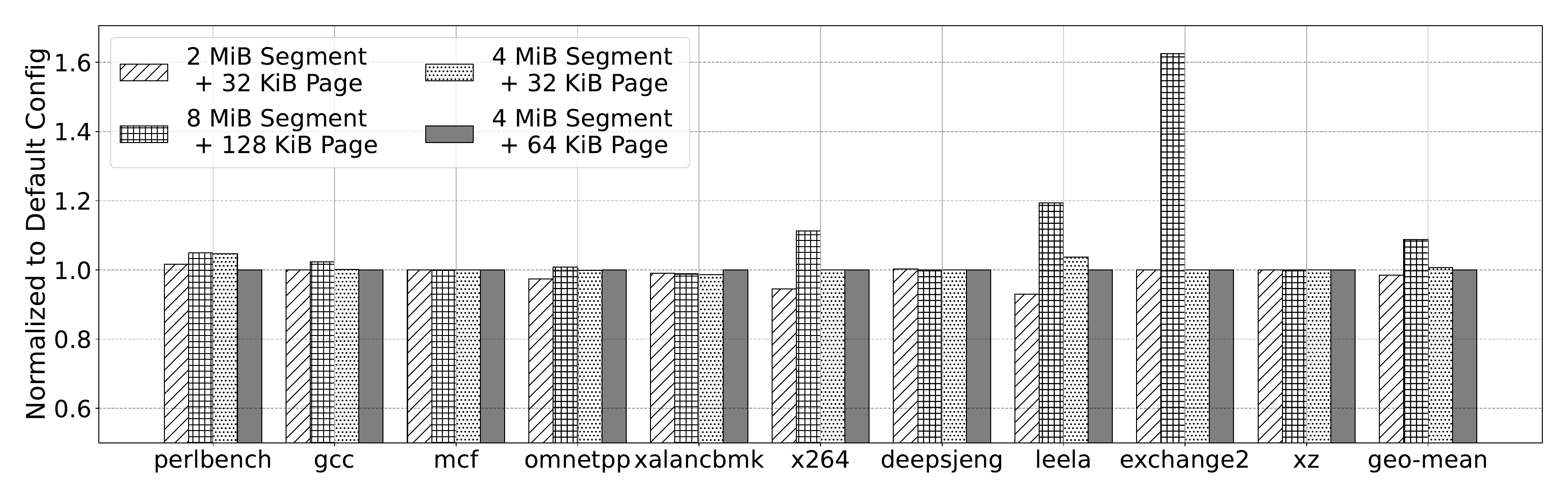}}
        \subcaption{Memory consumption on SPEC CPU2017 (smaller is better). }
	\label{fig_sensitivity_page_mem}
	\end{minipage} 
 \caption{Comparing the performance and memory consumption of \ExMalloc design choices under different page and segment size configurations. A 4 MiB segment size with 64 KiB small pages offers the best overall trade-off balance between performance and memory consumption, effectively accommodating diverse allocation patterns.  }
 \label{fig_sensitivity_page}
\end{figure}

\subsubsection{Sensitivity on segment and page size}
We initially set the segment size to 4 MiB based on empirical intuition and further validated this choice through a sensitivity study. 
Each segment contains 64 small pages; therefore, changing the segment size proportionally adjusts the page size. 
In this section, we evaluate four segment–page size configurations -- (4 MiB, 64 KiB), (4 MiB, 32 KiB), (8 MiB, 128 KiB), and (2 MiB, 32 KiB) -- to analyze their performance impact. \par

Fig.~\ref{fig_sensitivity_page} presents the performance comparison and memory consumption of these configurations on the SPEC CPU2017 workloads running on System A. 
Although the (8 MiB, 128 KiB) configuration provides a slight performance gain ($0.7\%$) over our chosen (4 MiB, 64 KiB) setting (Fig.~\ref{fig_sensitivity_page_perf}), it incurs substantially higher memory consumption (Fig.~\ref{fig_sensitivity_page_mem}).
Across the ten SPEC CPU2017 workloads (geometric mean), the (4 MiB, 64 KiB) configuration achieves an $8.7\%$ reduction in memory usage compared to (8 MiB, 128 KiB).
These results suggest that (4 MiB, 64 KiB) offers the best overall trade-off.

\begin{figure}[t]
	\begin{minipage}{\textwidth}
		\centerline{\includegraphics[width=0.95\columnwidth, trim = 9mm 9mm 28mm 8mm, page=1, clip=true]{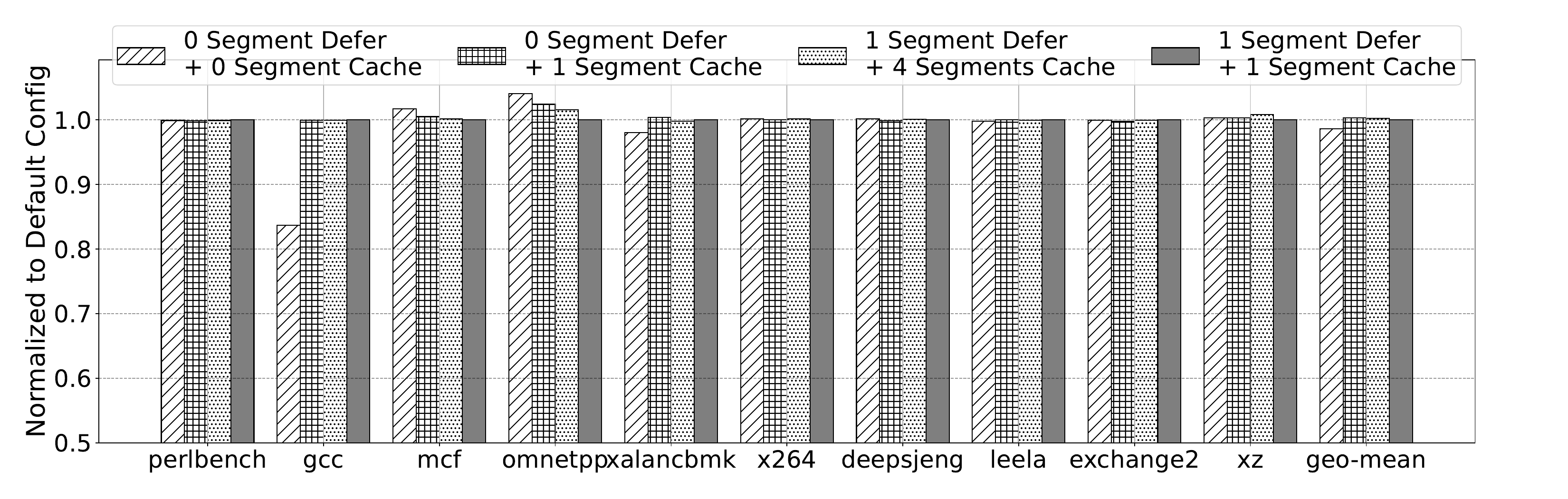}}
	    \subcaption{Performance on SPEC CPU2017 (larger is better). }
	    \label{fig_sensitivity_commit_perf}
	\end{minipage} 
 	\begin{minipage}{\textwidth}
		\centerline{\includegraphics[width=0.95\columnwidth, trim = 9mm 9mm 8mm 8mm, page=1, clip=true]{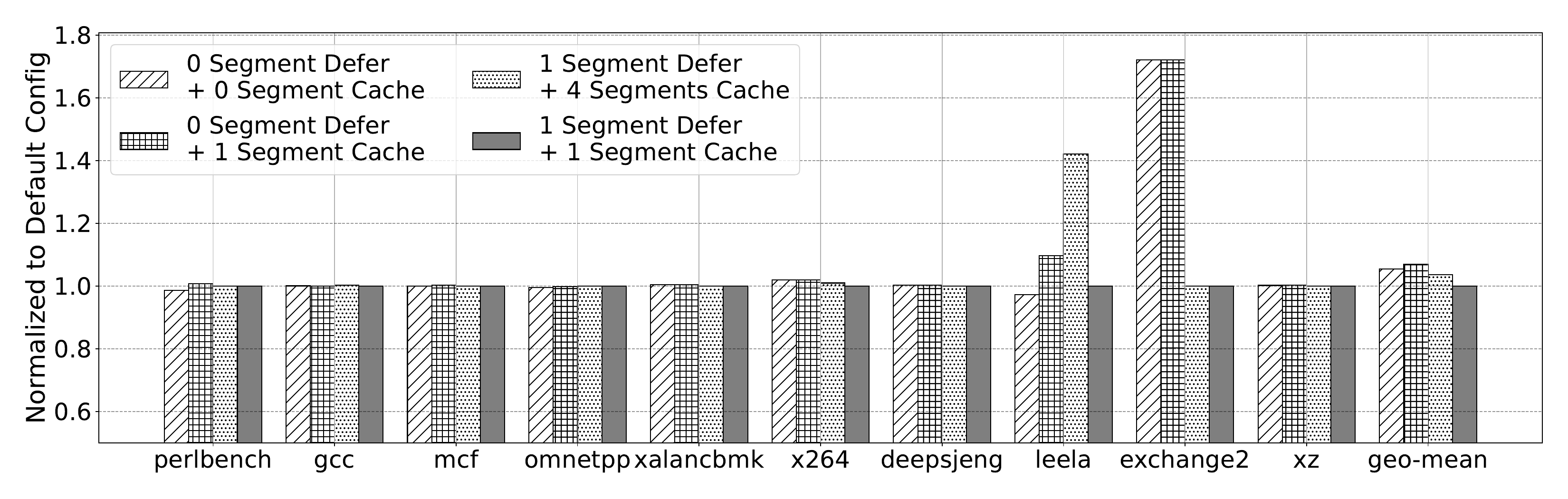}}
        \subcaption{Memory consumption on SPEC CPU2017 (smaller is better). }
	\label{fig_sensitivity_commit_mem}
	\end{minipage} 
 \caption{Comparing the performance and memory consumption of \ExMalloc design choices under different memory commitment and reclamation schemes. Having one deferred commitment segment and one cached segment for reclamation offers the best overall trade-off.  }
 \label{fig_sensitivity_commit}
\end{figure}

\subsubsection{Sensitivity on memory commitment and reclamation mechanisms}
\ExMalloc intentionally defers committing the first segment to the operating system and maintains one cached segment per page type during reallocation. 
This design eliminates the overhead of frequent $mmap()$ and $munmap()$ system calls, allowing rapid segment reuse and improving both allocation latency and memory efficiency.
To validate these design choices, we compare \ExMalloc against three alternative configurations: (a) no deferred commitment and no segment caching (0 Segment Defer + 0 Segment Cache), (b) no deferred commitment and one segment caching (0 Segment Defer + 1 Segment Cache), (c) one segment deferred commitment and four segments caching (1 Segment Defer + 4 Segments Cache), and our used configuration in \ExMalloc (d) one segment deferred commitment and one segment caching (1 Segment Defer + 1 Segment Cache). \par

Fig.~\ref{fig_sensitivity_page} shows the performance and memory consumption of these configurations on the SPEC CPU2017 workloads running on System A.
While the (0 Segment Defer + 1 Segment Cache) configuration delivers a slight performance improvement ($0.3\%$) over our selected (1 Segment Defer + 1 Segment Cache) setup (Fig.~\ref{fig_sensitivity_commit_perf}), it results in significantly higher memory consumption (Fig.~\ref{fig_sensitivity_commit_mem}).
Across the ten SPEC CPU2017 workloads (geometric mean), the (1 Segment Defer + 1 Segment Cache) configuration reduces memory usage by $7.0\%$ compared to (0 Segment Defer + 1 Segment Cache).
Overall, these results indicate that (1 Segment Defer + 1 Segment Cache) provides the best balance between performance and memory efficiency.

%% file: 06_conclusion.tex
Modern memory allocators are designed to support multi-threaded applications. To do so, these memory allocators often rely on complex metadata structures and control logic.
While this complexity enables them to handle diverse and dynamic allocation patterns in modern applications, it also introduces non-trivial overhead. 
In contrast, such overhead can be avoided in single-threaded contexts, which remain prevalent across a broad range of applications. \par

To address this gap, we propose \ExMalloc, a memory allocator designed specifically for single-threaded applications that retains a single-threaded design while integrating key principles from modern multi-threaded allocators.
\ExMalloc achieves a geometric mean speedup of $1.17\times$ over dlmalloc and $1.05\times$ over mimalloc, a modern multi-threaded allocator developed by Microsoft, on the SPEC CPU2017 benchmark suite. 
On redis-benchmark and mimalloc-bench, \ExMalloc provides $1.10\times$ and $1.93\times$ performance improvements over dlmalloc. 
Beyond performance, \ExMalloc also improves memory efficiency, achieving $6.2\%$, $0.1\%$, and $25.2\%$ memory consumption savings compared to mimalloc across SPEC CPU2017, redis-benchmark, and mimalloc-bench.
Given the continued use of single-threaded applications for their simplicity and efficiency, \ExMalloc presents a promising direction for future computer system design. \par